\documentclass[twocolumn]{aastex63}

\pdfoutput=1 %for arXiv submission
\usepackage{amsmath,amstext}
\usepackage[T1]{fontenc}
\usepackage{apjfonts} 
\usepackage[figure,figure*]{hypcap}

\newcommand {\gen}[1]{\texttt{Gen-#1}}

\usepackage{xcolor, fontawesome} % url, microtype
\definecolor{twitterblue}{RGB}{64,153,255}
\newcommand{\twitter}[1]{\href{https://twitter.com/#1}{\textcolor{twitterblue}{\faTwitter}\,\tt \textcolor{twitterblue}{@#1}}}

\shorttitle{EOS N-Body Simulations}
\shortauthors{Mulders, Ciesla, O'Brien, et al.}

\begin{document}

\title{Earths in Other Solar Systems N-body simulations: the Role of Orbital Damping in Reproducing the Kepler Planetary Systems}

\author[0000-0002-1078-9493]{Gijs D. Mulders}
\affiliation{Department of the Geophysical Sciences, University of Chicago \twitter{GijsMulders}}
\affiliation{Earths in Other Solar Systems Team, NASA Nexus for Exoplanet System Science}

\author{David~P.~O'Brien}
\affiliation{Planetary Science Institute, 1700 E.~Ft.~Lowell, Suite 106, Tucson, AZ 85719, USA}

\author{Fred J. Ciesla}
\affil{Department of the Geophysical Sciences, The University of Chicago, 5734 South Ellis Avenue, Chicago, IL 60637}
\affil{Earths in Other Solar Systems Team, NASA Nexus for Exoplanet System Science}

\author{D\'aniel Apai}
\affil{Department of Astronomy, The University of Arizona, Tucson, AZ 85721, USA}
\affil{Lunar and Planetary Laboratory, The University of Arizona, Tucson, AZ 85721, USA}
\affil{Earths in Other Solar Systems Team, NASA Nexus for Exoplanet System Science}

\author{Ilaria Pascucci}
\affil{Lunar and Planetary Laboratory, The University of Arizona, Tucson, AZ 85721, USA}
\affil{Earths in Other Solar Systems Team, NASA Nexus for Exoplanet System Science}

\begin{abstract}
The population of exoplanetary systems detected by Kepler provides opportunities to refine our understanding of planet formation. Unraveling the conditions needed to produce the observed exoplanet systems will allow us to make informed predictions as to where habitable worlds exist within the galaxy.
In this paper, we examine using N-body simulations how the properties of planetary systems are determined during the final stages of assembly, when planets accrete from embryos and planetesimals.  While accretion is a chaotic process, trends emerge allowing certain features of an ensemble of planetary systems to provide a memory of the initial distribution of solid mass around a star prior to accretion.  
We also use EPOS, the Exoplanet Population Observation Simulator, to account for detection biases and 
show that different accretion scenarios can be distinguished from observations of the Kepler systems.
We show that the period of the innermost planet, the ratio of orbital periods of adjacent planets, and masses of the planets are determined by the total mass and radial distribution of embryos and planetesimals at the beginning of accretion. In general, some amount of orbital damping, either via planetesimals or gas, during accretion is needed to match the whole population of exoplanets. 
Surprisingly, all simulated planetary systems have planets that are similar in size, showing that the "peas in a pod" pattern can be consistent with both a giant impact scenario and a planet migration scenario.
The inclusion of material at distances larger than what \textit{Kepler} observes ($>1$ au) 
has a profound impact on the observed planetary architectures, and thus on the formation and delivery of volatiles to possible habitable worlds.  
\end{abstract}


\section{Introduction}
Terrestrial planets form through the piece-wise collisional aggregation of solids spanning many orders of magnitude in size, beginning with micron-sized grains and continuing upward.  The final stages of growth involve the accretion of swarms of planetesimals and embryos, bodies measuring hundreds to thousands of kilometers across.  It is at these sizes where gravitational interactions lead to changes in the paths of these bodies around their central star, shifting them from the roughly circular orbits they develop within a gaseous protoplanetary disk, to more eccentric, crossing orbits,  that result in collisions and thus growth.  As protoplanets grow larger in this manner, the gravitational interactions work to space out the larger surviving bodies, isolating them from one another as the initial building blocks are largely cleared by further accretion, scattering out of the system, or collisions with the central star.  This leaves behind a fairly stable system of planets, with a small amount of leftover primordial bodies \citep[e.g.][]{raymond14}.

One possible outcome of this evolution is seen in the Solar System today, with planets orbiting at relatively large separations and with populations of small bodies within the asteroid belt, Kuiper Belt, and Oort Cloud.  How the Solar System achieved this structure has been the focus of a number of studies that used N-body codes to simulate the final stages of planetary growth, tracking how a distribution of embryos and planetesimals interact and grow over time \citep[e.g.][]{wetherill94,chambers98, Chambers2001Icar, Raymond2004Icar, raymond05, raymond06, raymond09, obrien06, OBrien2014Icar, Bond2010Icar, Bond2010ApJ, CarterBond2012bApJ, Fischer2014EPSL, walsh11, Quintana2007ApJ, Quintana2014ApJ, Izidoro2013ApJ, Izidoro2014ApJ}.  The success of these models in understanding the final stages of planet formation has largely been evaluated by their ability to reproduce the properties of the Solar System.  That is, key dynamical features (such as the mass of the asteroid belt, the angular momentum of the planets, their mass distribution, the low mass of Mars, and the timing of the last accretion event of the Earth) have been used to constrain the orbital properties of Jupiter and Saturn during terrestrial planet formation \citep{obrien06,raymond09}, the radial distribution of mass of the planetesimals and embryos \citep{raymond06}, or the extent to which Jupiter migrated while the solar nebula was still present \citep{walsh11}.

These studies have also found that similar initial conditions can yield very different final planetary systems \citep[e.g.][]{lissauer07,Fischer2014EPSL}.  The chaotic nature of the final assembly phase presents a challenge when comparing our Solar System to the model outcomes.  That is, it cannot be said with certainty whether our inability to simultaneously match all of the properties of the Solar System is due to the stochastic nature of the process and infinite number of possible outcomes that could be produced or if the initial conditions used in the models differ from those of reality.

While comparisons to the Solar System is a natural starting point to understand this stage of evolution, the discovery of thousands of exoplanetary systems has shown that our planetary system represents just one possible outcome of the  planet formation process \citep[e.g.][]{batalha14}.    The large number of systems discovered now allows us to identify general trends and relationships between key properties that arise during planetary assembly.  Thus, these systems provide a new opportunity to test our understanding of the final stages of planet formation by investigating how well our models reproduce general trends in observed exoplanet systems.  

Here we report on our study that explores how the key properties of planetary systems relate back to the initial conditions when planetary accretion begins and what conditions were necessary to produce the final architectures of the planetary systems found around sun-like stars.
Our goal is not to define or identify the best, single set of conditions that will reproduce the known planetary systems, but instead, to understand the connections and relationships between initial conditions that are responsible for defining the properties of the systems that we see today as a means of understanding formation pathways, and the bulk properties, for those planets \citep{apaiWP}.

Our approach here is to build a statistical framework in which we can model the outcomes of various planet formation scenarios and evaluate them against the known systems of exoplanets.   \citet{hansenmurray13} carried out an initial study with these goals, arguing that the known \textit{Kepler} systems could have formed if the final stages of assembly began with a massive disk of planetesimals concentrated within $\sim$1 au of their host stars. In particular, the distributions of orbital period and planet radius, as well as the relative frequencies of observed multi-planet systems produced in the models were consistent with those real systems that had been detected. Subsequent studies have corroborated and expanded on these results, focusing on the dynamical effects of residual gas in these simulations \citep{dawson15,dawson16} and explaining the apparent excess of single-transiting systems \citep{moriarty16}.

Currently, the number of Kepler planet candidates has grown to over four thousand \citep{thompson17} and the survey detection efficiency has been thoroughly characterized \citep[e.g.][]{burke15,christiansen16}, 
providing a more robust set of data to describe actual planetary properties to which model systems can be compared to.  
We explore a wide range of parameter space, and generate
 a large number of model planetary systems in order to understand the distribution of outcomes possible for a given range of parameters. We then evaluate how well the distribution of planetary properties produced by these initial conditions compares to  the Kepler dataset using the Exoplanet Population Observation Simulator \citep[\texttt{epos}\footnote{\url{https://github.com/GijsMulders/epos}},][]{epos1,epos2}.  Again, we do not expect to identify a single set of parameters or scenarios that describes the formation of all known planetary systems; instead, we are evaluating whether particular signposts of the imprints of the initial conditions and process of planetary accretion can be identified within the data sets. Success would then inform us of the formation histories of these planets. In the next section, we describe the various sets of initial conditions and dynamical effects we considered in our simulations.  The results of our simulations, which are part of the \texttt{Genesis} database of the \textit{Earths in Other Solar Systems} project\footnote{\url{http://eos-nexus.org/genesis-database}}, are presented in Section 3.  We compare the model results to the Kepler system using version 3.0 of the new \texttt{epos} tool in Section 4 to evaluate how well our models and assumptions about the final stages of planet formation match the real planetary systems that have been identified to date.  We conclude with a discussion and outline of future work.

\section{Dynamical Scenarios}

In all of our simulations, we follow previous work by assuming the distribution of solids around a solar mass central star follows a power law given by a surface density $\sigma = \sigma_o \left( \frac{r}{\mathrm{1~au}} \right)^{-p}$, where $\sigma_o$ is the surface density at a semimajor axis of $r=1$ au with index $p$. This prescription defines how much mass is available to build planets and how radially concentrated it is with respect to the star.  Some fraction of this mass, $f_{e}$, is assumed to be present in the form of embryos
which interact through gravitational interactions with all other bodies in the system.  The remaining fraction of solids, $f_{p}=1-f_{e}$, is taken to be present as planetesimals, which are smaller, less massive objects that interact gravitationally with the embryos but not with each other (a common computational simplification for these type of simulations).  All of these bodies are then taken to be distributed across an annulus ranging from an inner radius, $a_{min}$, to an outer radius $a_{max}$, beyond which it is assumed no other solids are present.

  In all cases, we assume that embryos are spaced an average of $\Delta$=7.5 mutual Hill radii apart, and thus have initial masses and semi-major axes defined by the relations in \citet{Kokubo2000Icar}.  For surface density slopes smaller than $p=2$, embryo mass increases with semi-major axis. Following \citet{hansenmurray13}, the embryos are placed on initial orbits that are nearly circular, with eccentricities randomly assigned values between 0 and 0.001, and orbital inclinations are randomly assigned from 0 to 2$^{\circ}$. 
 The orientations of the orbits and the initial location of each body on it is determined from a uniform distribution to define the final set of orbital parameters. The initial orbital parameters of the planetesimals are defined in a similar manner, although in their case the individual planetesimal mass is fixed and semimajor axes are chosen so that their distribution follows the surface density profile.  For the purposes of calculating the radii of the bodies of interest, we assume a density of $\rho$=3 g cm$^{-3}$.    The inner edge of the simulations is set to half the semi-major axis of the innermost body at the start of the simulations, and the timestep is set to 1/20 of the orbital period at the cutoff distance.  Bodies that pass interior to this cutoff distance are removed from the simulation.

\begin{figure}
\includegraphics[width=\linewidth]{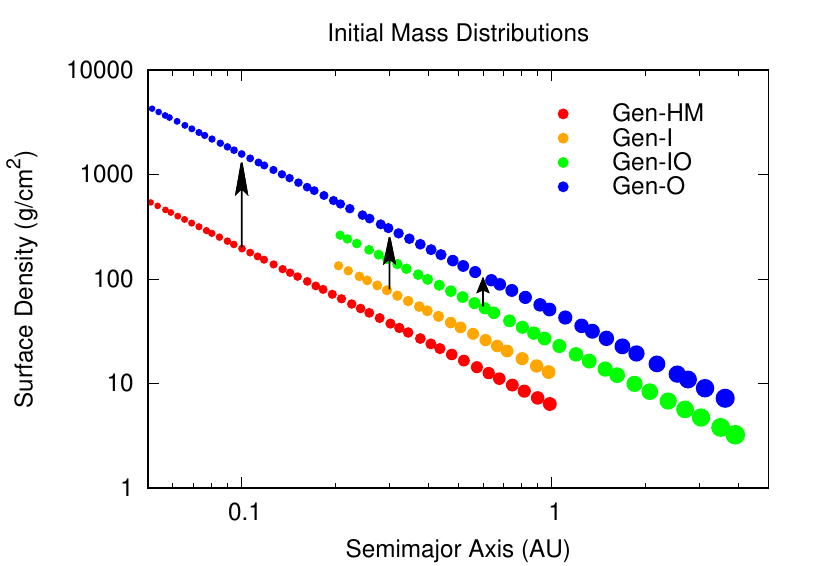}
\includegraphics[width=\linewidth]{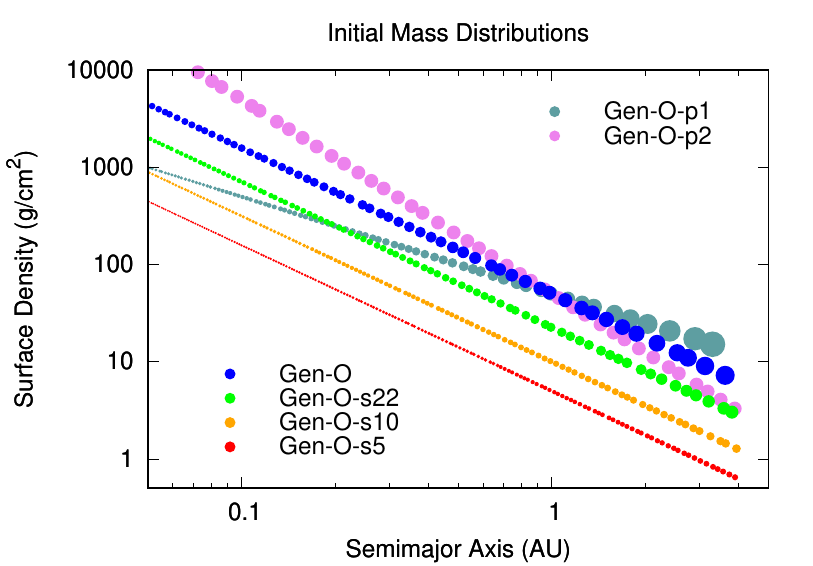}
\caption{Starting surface density distribution for simulations with different initial conditions. Symbol size is proportional to embryo radius (assuming a constant density). The top panel shows simulations with different inner and outer radii, vertically offset for clarity. The bottom panel shows simulations with different surface density normalization factors and exponents.
\label{f:initial}}
\end{figure}

Each of the simulations we perform is initialized by defining the parameters described above ($\sigma_{o}$, $p$, $f_{e}$, $a_{min}$, and $a_{max}$), and for each set of parameters many individual simulations are run with randomization of the orbital parameters as described above.  We refer to each of these as a \emph{Run Set}.  The parameter values for all of the runs we consider here are listed in Table \ref{t:sim}. Figure \ref{f:initial} also shows how the initial conditions compare across the various Run Sets, comparing and contrasting the initial radial distribution of the embryos and their masses.  Below we describe the broad categories of simulations that we performed.

%The orbital evolution of bodies in the gas-free simulations are integrated with the SWIFTER version of the SyMBA integrator \citep{Duncan1998AJ}, which handles close encounters and mergers\footnote{http://www.boulder.swri.edu/swifter/}.

\begin{deluxetable*}{l|ccccccccc}[t]
\tablecaption{Key parameters for the different sets of runs that we performed.  Values not shown here that were held constant across all runs are $M_{star}$ = 1 $\mathrm{M_{\odot}}$, the initial embryo spacing $\Delta$ = 7.5 mutual Hill radii, embryo and planetesimal density $\rho$ = 3 $\mathrm{g/cm^3}$, and initial eccentricity and inclination $e_o$ = 0.001 and $i_o$ = 2${}^o$. The parameter values that were varied from the base model \gen{HM} are denoted in bold face. 
\label{t:sim}}
\tablehead{\colhead{Run Name} & \colhead{$\sigma_o \ (\mathrm{g/cm^2})$} & \colhead{$p$} & \colhead{$a_{min}$ (au)} & \colhead{$a_{max}$ (au)} & \colhead{$a_{min,gas}$ (au)} & \colhead{$f_{emb}$} & \colhead{$N_{emb}$} & \colhead{$N_{ptsml}$} & \colhead{$N_{run}$}}
\startdata
\texttt{Gen-HM} &		50.00 &	1.50 &	0.05 &	1.00 &	n/a &	1.00 &	45 &	0  &	50 \\
\texttt{Gen-P} &		50.00 &	1.50 &	0.05 &	1.00 &	n/a &	\bf 0.75 &	51 &	1027 &	\bf 8 \\
\texttt{Gen-I} &		50.00 &	1.50 &	\bf 0.20 &	1.00 &	n/a &	1.00 &	20 &	0 &	50 \\
\texttt{Gen-P-I} &	    50.00 &	1.50 &	\bf 0.20 &	1.00 &	n/a &	\bf 0.75 &	23 &	733 &	\bf 8 \\
\texttt{Gen-IO} &		50.00 &	1.50 &	\bf 0.20 &	\bf 4.00 &	n/a &	1.00 &	32 &	0 &	50 \\
\texttt{Gen-P-IO} &		50.00 &	1.50 &	\bf 0.20 &	\bf 4.00 &	n/a &	\bf 0.75 &	36 &	2055 &	\bf 8 \\
\texttt{Gen-O} &		50.00 &	1.50 &	0.05 &	\bf 4.00 &	n/a &	1.00 &	56 &	0 &	50 \\
\texttt{Gen-O-p1} &		50.00 &	\bf 1.00 &	0.05 &	\bf 4.00 &	n/a &	1.00 &	79 &	0 &	50 \\
\texttt{Gen-O-p2} &		50.00 &	\bf 2.00 &	0.05 &	\bf 4.00 &	n/a &	1.00 &	44 &	0 &	50 \\
\texttt{Gen-O-s22} &		\bf 22.50 &	1.50 &	0.05 &	\bf 4.00 &	n/a &	1.00 &	84 &	0 &	50 \\
\texttt{Gen-O-s10} &		\bf 10.00 &	1.50 &	0.05 &	\bf 4.00 &	n/a &	1.00 &	126 &	0 &	50 \\
\texttt{Gen-O-s5} &		\bf 5.00 &	1.50 &	0.05 &	\bf 4.00 &	n/a &	1.00 &	178 &	0 &	50 \\
\texttt{Gen-M-s50} &	\bf 50.00 &	1.50 &	0.05 &	\bf 10.00 &	\bf 0.05 &	1.00 &	62 &	0 &	50 \\
\texttt{Gen-M-s22} &	\bf 22.50 &	1.50 &	0.05 &	\bf 10.00 &	\bf 0.05 &	1.00 &	93 &	0 &	50 \\
\texttt{Gen-M-s10} &	\bf 10.00 &	1.50 &	0.05 &	\bf 10.00 &	\bf 0.05 &	1.00 &	139 &	0 &	50 \\
\enddata
\end{deluxetable*}

\subsection{Embryo-only}
We performed several sets of simulations that consisted only of massive embryos ($f_{e}$=1) based on the initial conditions used in \citet{hansenmurray13} that were found to reproduce the broad properties of planetary systems in terms of planet radii, orbital periods, and multi-planet frequency in the \textit{Kepler} catalog at the time.  The nominal case (which we refer to as \gen{HM}) is chosen to approximately reproduce their simulations, with surface mass density profile exponent $p$=3/2, normalization factor $\sigma_o = 50 \ \mathrm{g/cm^{2}}$ at 1 au, and embryos distributed between $a_{min}$ of 0.05 au and $a_{max}$ of 1 au.   This gives 45 total embryos, with a total mass of 19 M$_{\bigoplus}$ (in comparison, \citet{hansenmurray13} had 32-38 embryos and 20 M$_{\bigoplus}$, likely due to a larger spacing between embryos). In addition to the nominal case, we systematically explored a number of other initial conditions, varying the radial range over which building blocks were distributed ($a_{min}$ - $a_{max}$, simulations \gen{I}, \gen{IO}, and \gen{O}), the normalized surface density $\sigma_o$ at 1 au (\gen{O-s22}, \gen{O-s10}, \gen{O-s5}), and the exponent $p$ of the power law that describes how the mass is distributed with radius (\gen{O-p1}, \gen{O-p2}. The surface density is a factor of five higher than what is needed to produce the terrestrial planets in the solar system \citep[e.g.][]{raymond06}. Most simulations were carried out for 10 Myr, though we also ran many for longer periods of time (up to 80 Myr) to explore how planetary system structure varies over these longer timescales.  The orbital evolution simulations were performed using the SWIFTER version of the SyMBA integrator \citep{Duncan1998AJ}, which handles close encounters and mergers\footnote{http://www.boulder.swri.edu/swifter/}. 
We assume that all collisions result in perfect mergers. Including fragmentation in the simulations would prolong the time-scale for planetary accretion, but generally results in similar planet planetary system properties as most collisional products are re-accreted on the same planet \citep{chambers13}.

\subsection{Planetesimals and Embryos}

We also performed sets of simulations with a population of planetesimals distributed among the embryos (\gen{P}, \gen{P-I}, and \gen{P-IO}).  These small bodies have been shown to damp down the inclinations and eccentricities of forming planets through dynamical friction \citep[e.g.,][]{Wetherill1993Icar,obrien06}.  In these cases, we used the same total mass of solid materials as for our embryo-only cases, but distributed it with 75\% being present as embryos initially ($f_{emb} = 0.75$) and the remaining 25\% initially being contained in planetesimals.  In our \gen{P} case, which follows our nominal model, the planetesimal mass is distributed as 1000 distinct bodies, giving an individual planetesimal mass that is $\sim$1/20 of the smallest embryo ($\sim$0.005$M_{\oplus}$).  The individual planetesimal mass in \gen{P-I} and \gen{P-IO} is similarly calculated as 1/20 of the mass of the smallest embryos in those simulations, with the number of such bodies varying according to the total mass in the simulation (see Table 1). We again used the SWIFTER version of the SyMBA integrator for these calculations.

\subsection{Planet migration}
A key issue that can also affect the final architecture of planetary systems is how much of the final stage of planetary assembly occurs while the gas from the protoplanetary disk is still present.  To investigate the effects that nebular gas has on the emerging planetary systems,  we performed several sets of simulations where we follow the dynamical evolution of embryos while they are embedded in a gaseous protoplanetary disk, leading to radial migration and damping of their eccentricities and inclinations over time.  
In these cases, we use a modified version of the Mercury integrator \citep{Chambers1999MNRAS} developed by \cite{izidoro17}.  This code incorporates the 1-D time dependent density and temperature structure model of a viscous disk \citep{Bitsch2015AA}, which was derived from fits to a 2-D numerical hydrodynamical and radiative transfer simulation. The dominant form of migration for embryo-sized bodies is Type-1 migration, which is modeled as in \citet{Paardekooper2010MNRAS, Paardekooper2011MNRAS}, using methods developed by \citet{Cresswell2008AA}, \citet{Coleman2014MNRAS}, \citet{Fendyke2014MNRAS}, and \citet{Papaloizou2000MNRAS}.  Tidal eccentricity and inclination damping of the embryos is also included \citep{Papaloizou2000MNRAS, Tanaka2004ApJ, Cresswell2006AA, Cresswell2008AA}.  The disk has a lifetime of 5 Myr, and the disk structure is updated every 500 years with a resolution of 0.05 au. 
The disk dispersal is described by an exponential decrease in disk surface density with a e-folding timescale of 0.01 Myr, see \cite{izidoro17} for details.

In the disk models, the relevant parameters are the metallicity (dust-to-gas ratio) of 0.01, adiabatic index $\gamma$=1.4 and alpha viscosity $\alpha = \mathrm{5.4 \times 10^{-3}}$.  Together, they define the rate of mass and angular momentum transport through the disk, and the resulting temperature profile as a result of accretional heating and stellar irradiation.  While the code is able to account for stochastic effects from MRI-driven turbulence, it is not included in the current simulations as \cite{izidoro17} found that it did not have a significant effect.  The inner edge of the gas disk is set to 0.05 au (the same as the inner edge of the initial embryo distribution in \gen{HM} and \gen{O}), and the surface density near the inner edge is smoothed off using a hyperbolic tangent \citep[][Eq. 4]{izidoro17}.

The specific setup of the gas disk model is one where planets migrate inward through a continuous disk and get trapped near the inner edge of the gas disk and in orbital resonances with each other. This scenario is consistent with the observed peak in the location of planetary system inner edges at $\sim{}10$ days \citep{epos1,carrera19}. We do not explore alternate scenarios for trapping of planets farther out on the disk, for example if the inner disk structure is modified by MHD disk winds \citep{ogihara15}.

\begin{figure*}
\includegraphics[width=\linewidth]{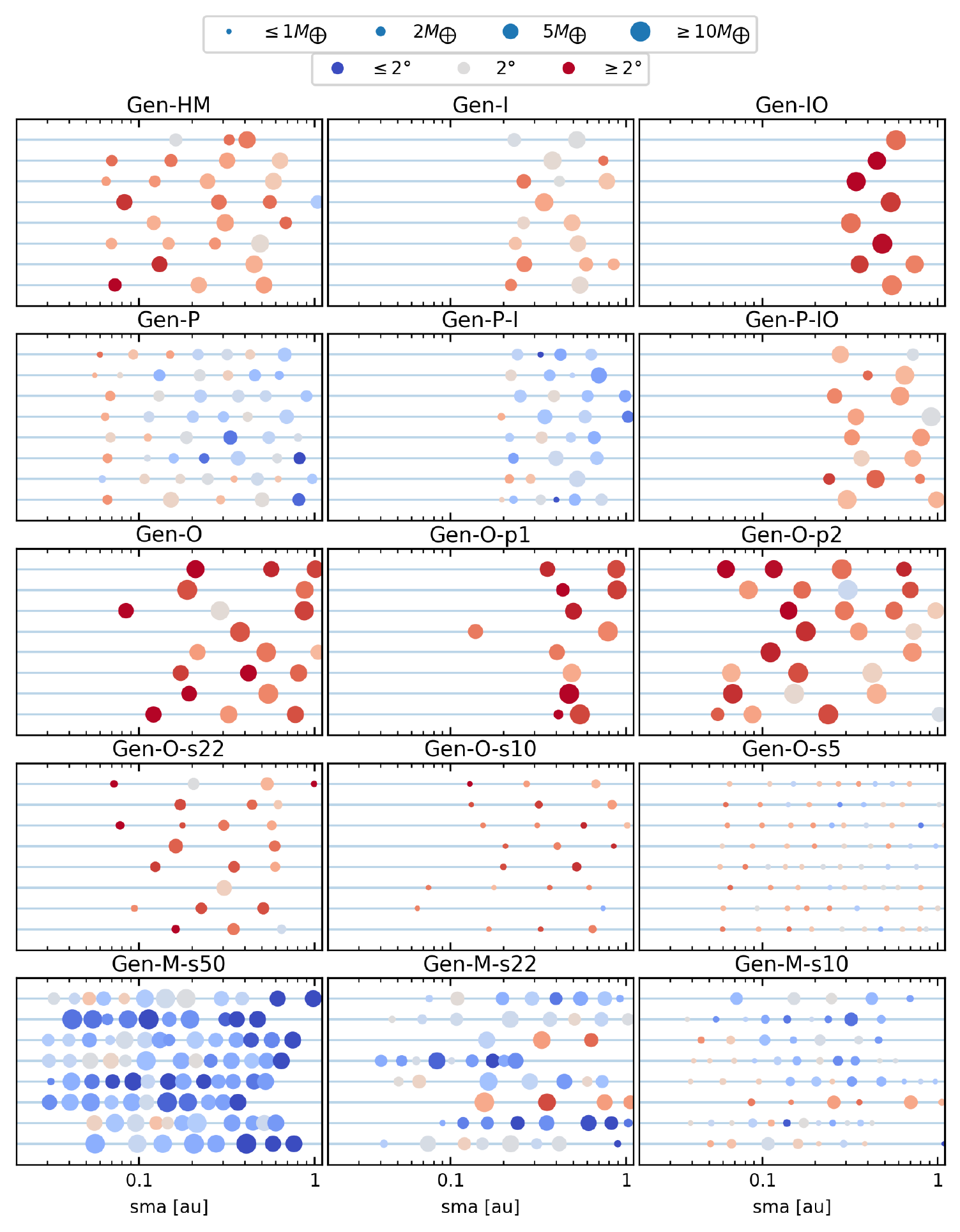}
\caption{Planet properties and architectures of systems formed in each run set. The symbol size corresponds to planet mass. The symbol color corresponds to the orbital inclination, where red denotes inclinations larger than the $2\degr$ inferred from \textit{Kepler} and  blue indicates lower inclinations.
Only 8 systems per run are shown, the full set of figures and data will be available as an online supplement. 
%\dave{Seems like it could all go in supplemental material along with the publication}.
\label{f:individual}}
\end{figure*}

\section{Architecture of Simulated Planetary Systems}
Our primary goal in running the suite of simulations described above is to understand how the initial conditions of the final stage of planet formation impact the properties of the planetary systems that formed.  The physical framework for the final stages of planetary accretion is that as the initial swarms of planetesimals and embryos around a star begin to gravitationally interact, they perturb one another's orbits, leading to orbital crossings and collisions.  Early on, these collisions are frequent as the bodies begin tightly packed and low in mass, yielding many accretionary events.  These events are more frequent close to the star, as the shorter orbital periods there allow for a greater number of orbital crossing events in a given time, increasing the likelihood of accretionary impacts.  As bodies grow more massive, they begin to dominate their local annulus, either accreting other bodies or scattering them away from their neighborhood.  As bodies begin to dominate their local annulus, their gravitational interactions with neighbors serve to keep these objects on orbits that are more widely spaced and dynamically stable.  Collisions still occur as orbital perturbations can grow, but such events typically become rarer as the number of bodies in orbit around the star decreases.  

While one might expect systems that begin with similar initial conditions to yield planetary systems with similar properties, it is important to account for the chaotic nature of this stage of evolution.
That is, while planetesimals and embryos may begin this stage with identical orbital properties, the outcomes are very sensitive to where along their orbits the bodies begin (something that is randomized between each particular simulation). Thus for each Run Set listed in Table \ref{t:sim}, we performed multiple simulations in order to provide a statistical measure of the distribution of possible outcomes.  In the embryo-only cases, we typically performed 50 realizations of the simulations; the runs with planetesimals were much more computationally demanding, and thus we typically performed 8 simulations for those cases. 

\begin{deluxetable}{l|cccccc}[t]
\tabletypesize{\scriptsize}
\tablecaption{Simulated System Properties inside 1 au.
Each column shows the median value and 1-sigma ranges in each set of simulations of planet mass, $M_p$, mass ratio, $\mathcal{M}$,  mutual inclination, $i$, period of innermost planet, $P_\text{inner}$, period ratio, $\mathcal{P}$, and number of planets per system $\overline{N}$.
\label{t:numbers}}
\tablehead{
\colhead{Run Name} 
&\colhead{$\overline{M}_p$ ($M_{\oplus}$)} 
&\colhead{$\overline{\mathcal{M}}$}
&\colhead{$\overline{P}_\text{inner}$ (days)}
&\colhead{$\overline{\mathcal{P}}$}
&\colhead{$\overline{i}$ ($\deg$)}
&\colhead{$\overline{N}$}
}
\startdata
\texttt{Gen-HM} & $3.7^{+2.1}_{-1.4}$ & $1.2^{+0.94}_{-0.51}$ & $7.5^{+8.4}_{-1.5}$ & $2.7^{+1.1}_{-0.68}$ & $5.2^{+4.6}_{-3.2}$ & $4^{+1}_{-1}$ \\
\texttt{Gen-I} & $4.2^{+1.9}_{-1.4}$ & $0.9^{+0.98}_{-0.32}$ & $49^{+22}_{-7.3}$ & $2.6^{+0.89}_{-0.6}$ & $5^{+3.8}_{-2.7}$ & $2^{+1}_{-0}$ \\
\texttt{Gen-IO} & $8.9^{+2.3}_{-2.6}$ & $0.78^{+0.61}_{-0.21}$ & $78^{+11}_{-17}$ & $4.2^{+0.66}_{-1.4}$ & $14^{+9.7}_{-5.6}$ & $1^{+1}_{-0}$ \\
\texttt{Gen-P} & $2.2^{+1.1}_{-0.59}$ & $1.1^{+0.8}_{-0.41}$ & $6.1^{+0.45}_{-0.69}$ & $2^{+0.27}_{-0.33}$ & $1.8^{+2.3}_{-0.85}$ & $6.5^{+0.5}_{-0.5}$ \\
\texttt{Gen-P-I} & $2.8^{+0.66}_{-1.3}$ & $1^{+1.8}_{-0.28}$ & $38^{+5.3}_{-5.2}$ & $1.8^{+0.46}_{-0.29}$ & $1.1^{+1.3}_{-0.54}$ & $4^{+0}_{-0.89}$ \\
\texttt{Gen-P-IO} & $5.9^{+2.4}_{-3}$ & $1.3^{+1.3}_{-0.76}$ & $65^{+16}_{-16}$ & $3.6^{+0.71}_{-1.2}$ & $5.6^{+3.6}_{-2}$ & $2^{+0}_{-0}$ \\
\texttt{Gen-O} & $7.3^{+3.3}_{-2.4}$ & $1^{+0.5}_{-0.44}$ & $30^{+21}_{-13}$ & $4.4^{+4.7}_{-1.7}$ & $13^{+8.9}_{-6.2}$ & $2^{+1}_{-0.23}$ \\
\texttt{Gen-O-p1} & $7.1^{+4.5}_{-2.7}$ & $1.7^{+1}_{-1}$ & $54^{+49}_{-29}$ & $5.8^{+4.1}_{-3}$ & $16^{+14}_{-8.1}$ & $1^{+1}_{-0}$ \\
\texttt{Gen-O-p2} & $9.5^{+3.6}_{-2.4}$ & $1^{+0.55}_{-0.36}$ & $6.9^{+6.7}_{-1.3}$ & $3.4^{+1.5}_{-0.75}$ & $9.1^{+7.1}_{-5.5}$ & $3^{+1}_{-0}$ \\
\texttt{Gen-O-s22} & $2.2^{+1}_{-0.81}$ & $1.2^{+0.66}_{-0.54}$ & $11^{+13}_{-3.5}$ & $3.1^{+1.2}_{-0.88}$ & $9.2^{+7.5}_{-5.1}$ & $3.5^{+0.5}_{-0.5}$ \\
\texttt{Gen-O-s10} & $1^{+0.46}_{-0.34}$ & $1.3^{+0.6}_{-0.43}$ & $17^{+12}_{-9.4}$ & $3^{+1.3}_{-0.65}$ & $8.6^{+7.9}_{-3.9}$ & $3^{+1}_{-0}$ \\
\texttt{Gen-O-s5} & $0.17^{+0.094}_{-0.054}$ & $1^{+0.77}_{-0.43}$ & $5.5^{+0.5}_{-0.47}$ & $1.6^{+0.22}_{-0.2}$ & $3.4^{+2.7}_{-1.9}$ & $10^{+0.7}_{-1}$ \\
\texttt{Gen-M-s50} & $6.2^{+2.9}_{-2}$ & $1.1^{+0.45}_{-0.54}$ & $2.9^{+1.9}_{-0.92}$ & $1.5^{+0.17}_{-0.17}$ & $0.13^{+0.79}_{-0.12}$ & $10^{+2}_{-2.4}$ \\
\texttt{Gen-M-s22} & $2.3^{+2.6}_{-1.1}$ & $1.1^{+0.96}_{-0.58}$ & $3.3^{+9}_{-1.2}$ & $1.5^{+0.75}_{-0.25}$ & $0.31^{+2.5}_{-0.3}$ & $8.5^{+4.5}_{-4.5}$ \\
\texttt{Gen-M-s10} & $0.99^{+0.89}_{-0.5}$ & $0.99^{+0.72}_{-0.46}$ & $3^{+1.5}_{-0.81}$ & $1.4^{+0.39}_{-0.21}$ & $0.59^{+1.6}_{-0.55}$ & $10^{+7.5}_{-3}$ \\
\hline
\enddata
\end{deluxetable}

\begin{figure*}
\includegraphics[width=0.9\linewidth]{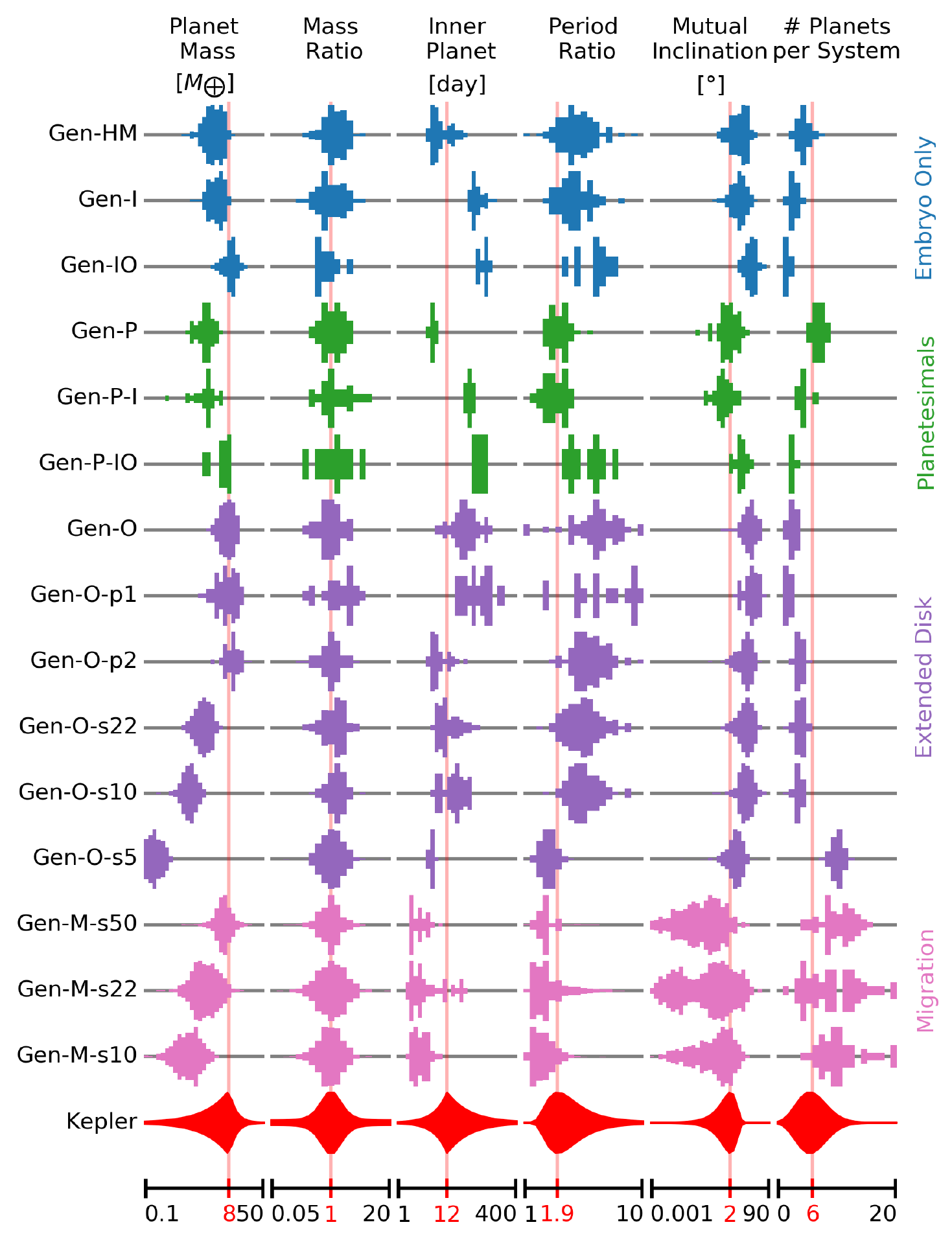}
\caption{Distribution of planetary system properties inside 1 au averaged across all (8 or 50) runs within one set of simulations. 
The simulations are color-coded by accretion scenario: embryo-only (blue); planetesimals (green); extended embryo/planetesimal disk (purple); and migration (pink); The de-biased distributions derived from \textit{Kepler} are shown in red, with a typical value of the distribution indicated across rows. The planet mass distribution is taken from \cite{pascucci18}, while the distributions of planet inclination, period ratio of adjacent planets, and inner planet location are those from \cite{epos1}. The number of planets per system is shown as a poisson distribution with 6 planets per system.
The bottom row shows the scale bar on a logarithmic stretch except for the number of planets per system which is shown on a linear scale. 
The peak of each distribution is normalized to span the entire y axis.
\label{f:tablehist}}
\end{figure*}

Figure \ref{f:individual} shows a sampling of the realizations produced within each Run Set.  While variations in outcomes within a given Run Set are readily seen, it is clear that these differences are small compared to those between the different Run Sets we considered.  This indicates that despite the stochastic nature of this stage of evolution, it is possible to identify signatures of the conditions under which planetary accretion took place.

 Ultimately, we wish to use the relation between input parameters and realizations of planetary systems to inform the formation histories of the planetary systems found by \emph{Kepler}.  In order to do that, we must identify key diagnostics of the \emph{Kepler} systems that can be compared to our models.  Given the limited data we have on real systems, we have identified a set of broad, quantifiable parameters that describe the properties of these systems.  Specifically, we use
 the median planet mass ($\overline{M}_p$), the mass ratio of adjacent planets ($\overline{\mathcal{M}}$), the median mutual inclination with respect to the invariable plane of the system, ($\overline{i}$), the median location of the innermost planet in the system, ($\overline{P}_\text{inner}$), the median orbital period ratio of adjacent planets, ($\overline{\mathcal{P}}$), and the median number of planets per system, $\overline{N}$.  It is important to note that these parameters focus on the properties of the planetary \emph{system}; that is we focus on the ensemble collection of planets, rather than the specific properties of individual planets, as these will likely inform us about how the planets interacted with one another during planet formation. 

In order to ensure we are making the best comparisons between the models and the planetary systems, we focus on those planets in our simulations that are contained inside of 1 au.  Because the detection efficiency of \emph{Kepler} drops significantly with distance from the star, few planets are known to exist outside of that distance, making it difficult to make robust and meaningful comparisons.  Note that we still allow planets to form in that region in our simulations; we just limit our statistics to those located closer to the star.
Table \ref{t:numbers} summarizes the median values and 1 $sigma$ ranges for each property for the Run Sets we considered while Figure \ref{f:tablehist} provides a visual comparison of the outcomes, providing not only the characteristic values, but also their distributions.

The parameter distributions have mostly distinct shapes between run sets, with the exception of the mass ratio distributions which are more similar across all run sets. In order to quantify the similarities between run sets we calculate the probability that the model distributions are drawn from the same underlying distribution with a KS test. Table \ref{t:ks} shows a set of comparions between all run sets and the base model, \gen{HM}. 
The distributions of most parameters ($M_p$, $P_\text{inner}$, $\mathcal{P}$, $i$, and $N$) are distinguishable form the base set at high confidence ($p<0.01$). The exception is the mass ratio between adjacent planets, $\mathcal{M}$, which are statistically indistinguishable from the base set for the majority of run sets. We find the same trend when doing these comparisons between all possible combination of run sets: only the planet mass ratio distributions are often similar between models.  
None of the run sets are indistinguishable in all six parameter distributions, indicating that the variations in initial conditions indeed result in real differences in planetary system architectures that can be traced by these diagnostics. 

\begin{deluxetable}{l|cccccc}[t]
\tabletypesize{\scriptsize}
\tablecaption{
Similarities and differences between run sets.
This table shows the KS-probability that the distribution of system properties (Fig. \ref{f:tablehist}) are drawn from the same distribution as the base set, \gen{HM}. The table headers are the same as those in Table \ref{t:numbers}. Only the mass ratio distributions are mostly consistent between the different run sets.
\label{t:ks}}
\tablehead{
\colhead{Run Name} 
&\colhead{${M}_p$ ($M_{\oplus}$)} 
&\colhead{${\mathcal{M}}$}
&\colhead{${P}_\text{inner}$ (days)}
&\colhead{${\mathcal{P}}$}
&\colhead{${i}$ ($\deg$)}
&\colhead{${N}$}
}
\startdata
\texttt{Gen-I} & 0.038 & 0.008 & $<$0.001 & 0.252 & 0.477 & $<$0.001\\
\texttt{Gen-IO} & $<$0.001 & 0.104 & $<$0.001 & $<$0.001 & $<$0.001 & $<$0.001\\
\texttt{Gen-P} & $<$0.001 & 0.559 & 0.001 & $<$0.001 & $<$0.001 & $<$0.001\\
\texttt{Gen-P-I} & $<$0.001 & 0.280 & $<$0.001 & $<$0.001 & $<$0.001 & 0.997\\
\texttt{Gen-P-IO} & 0.001 & 0.763 & $<$0.001 & 0.183 & 0.307 & $<$0.001\\
\texttt{Gen-O} & $<$0.001 & 0.067 & $<$0.001 & $<$0.001 & $<$0.001 & $<$0.001\\
\texttt{Gen-O-p1} & $<$0.001 & 0.028 & $<$0.001 & $<$0.001 & $<$0.001 & $<$0.001\\
\texttt{Gen-O-p2} & $<$0.001 & 0.022 & 0.481 & $<$0.001 & $<$0.001 & 0.055\\
\texttt{Gen-O-s22} & $<$0.001 & 0.374 & $<$0.001 & 0.025 & $<$0.001 & 0.272\\
\texttt{Gen-O-s10} & $<$0.001 & 0.132 & $<$0.001 & 0.001 & $<$0.001 & $<$0.001\\
\texttt{Gen-O-s5} & $<$0.001 & 0.062 & $<$0.001 & $<$0.001 & $<$0.001 & $<$0.001\\
\texttt{Gen-M-s50} & $<$0.001 & $<$0.001 & $<$0.001 & $<$0.001 & $<$0.001 & $<$0.001\\
\texttt{Gen-M-s22} & $<$0.001 & 0.005 & $<$0.001 & $<$0.001 & $<$0.001 & $<$0.001\\
\texttt{Gen-M-s10} & $<$0.001 & $<$0.001 & $<$0.001 & $<$0.001 & $<$0.001 & $<$0.001\\
\hline
\enddata
\end{deluxetable}

In Figure \ref{f:tablehist}, we also show the approximate values and derived distributions for the \emph{Kepler} exoplanets for reference. As we outline below, comparisons between the model realizations and the \emph{Kepler} systems must be done with caution, and we outline a robust methodology in the next section.  Before linking the properties of \emph{Kepler} systems to particular formation histories, it is necessary to understand how the values of various properties are related to the initial conditions of the runs.  Here we describe the effects of changes in the initial conditions on the realizations of planetary systems to provide a framework for interpreting the \emph{Kepler} properties.

\subsection{Effects of Including Planetesimals}\label{s:model:plts}

Planetesimals in our models represent the leftover building blocks of planetary embryos, the more massive bodies that are the primary gravitational perturbers during the final stages of planetary accretion.  Their incorporation into embryos is not expected to be 100\% efficient, and as such, a number of them are expected to be present while planets form.  The effects of planetesimals are seen when comparing the realizations from Run Set pairs \gen{HM} \& \gen{P}, \gen{I} \& \gen{P-I}, and \gen{IO} \& \gen{P-IO}; each of these simulation pairs began with the same amount of mass present over the same radial distribution, but the latter had 25\% of their mass contained in planetesimals and 75\% in embryos, while the former was entirely embryos.  

The primary effect of planetesimals is to provide a source of dynamical friction, where their orbits are preferentially excited to high eccentricities and inclinations while embryos and growing planets remain on more circular and roughly coplanar orbits.
Inspection of Figures \ref{f:individual} and \ref{f:tablehist}, as well as the numbers reported in Table \ref{t:numbers} show that the inclinations of the planets formed in the embryo-only simulations are significantly higher than their planetesimal-containing pairs.  

Another consequence of planetesimals being present during planetary accretion is that the number of planets that form in a given system increases.  This effect is related to dynamical friction; as the embryos in the system maintain less-excited orbits, their eccentricities and inclinations are damped, limiting the semi-major axis range over which embros can collide. As a result, the feeding zones are narrower, leaving a larger number of bodies behind.  This same process is also responsible for the lower masses of the planets in those cases where planetesimals were present: narrower feeding zones means fewer mass-adding events for any given planet (Figure \ref{f:final}a).

\subsection{Effects of Inner Radius}\label{s:model:inner}

The inner radius of our simulation defines the smallest orbital semi-major axis at which solid materials were available to form planets. 
 In simulations for the accretion of the Solar System, this inner edge is often set at $\sim$0.5 au for computational purposes as closer-in orbits require finer time resolutions to resolve \citep[e.g.][]{chambers98,chambers2004,obrien06,raymond09}.  Thus, these simulations largely focus on the accretion of planets outside of this location.  However, given that planetary systems discovered by \textit{Kepler} contain planets located much closer to their parent stars, this requires examining the details of planet formation at these very low orbital periods.

\begin{figure*}
\includegraphics[width=\linewidth]{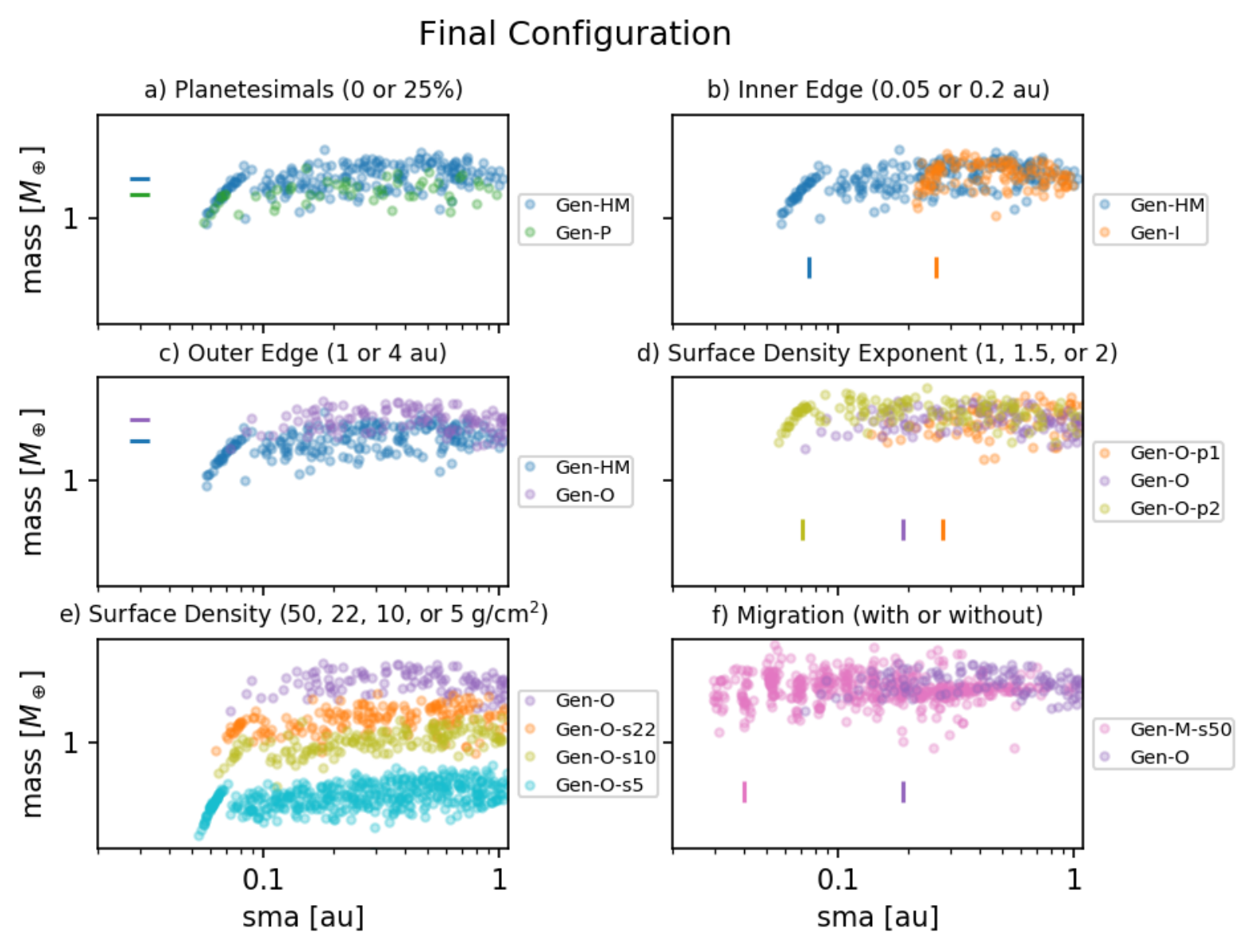}
\caption{Simulated planet mass and semi-major axes distributions under the influence of varying one specific parameter in the initial conditions. 
The horizontal dashes indicate the mean mass of planets within 1 au whenever that varies significantly between simulations. The vertical dashes indicate the location of the innermost planet per system if that location varies between simulations. 
From left to right, top to bottom, the specific parameters/processes varied are: 
a) The inclusion of planetesimals, which affects planet mass (\S \ref{s:model:plts}); 
b) The location of the disk inner edge, which affects the period of the innermost planet in each system (\S \ref{s:model:inner}); 
c) The location of the disk outer edge, which affects the planet mass (\S \ref{s:model:outer}); 
d) The slope of the radial surface density distribution, which affects the period of the innermost planet in each system (\S \ref{s:model:sdp}); 
e) The normalization of the surface density, which affects planet mass (\S \ref{s:model:sdp}); 
f) enabling gas-driven migration and orbital damping, which affects the period of the innermost planet in each system (\S \ref{s:model:gas}).
\label{f:final}}
\end{figure*}

Run Set pairs \gen{HM} \& \gen{I}, \gen{HM} \& \gen{P-I}, and \gen{O} \& \gen{IO} have identical properties to one another but vary the inner edge from 0.05 au (first simulation in pair) to 0.2 au (second). In all cases, the location of the innermost planet correlates with the inner edge of the disk, with those disks extending to 0.05 au yielding innermost planets with lower periods as a result of raw materials being present closer-in for planets to form from (Figure \ref{f:final}b). As a result, the number of planets formed within 1 au is larger for smaller inner disk edges.

Variations in planet masses, inclinations, and period ratios are relatively minor among the Run Set pairs (Table \ref{t:numbers}). In particular, the masses of planets outside of 0.2 au are not affected by the location of the disk inner edge (Figure \ref{f:final}b)

\subsection{Effects of Outer Radius}\label{s:model:outer}

With most known extrasolar planets, particularly those in the Kepler database, found in low period orbits, many simulations for their formation have focused on the assembly of local materials inside of 1 au \citep{hansenmurray12,hansenmurray13,dawson15,dawson16}.  However, given the decreased detection probability further from the star, it remains uncertain what the distribution of mass at larger distances may be.  To investigate the effects of mass at larger distances, we performed simulations where the outer edge of the solids extended to 4 au.  The outer edge of where mass may be located is likely set by a combination of the radial extent of the disk as well as how solids migrate prior to being incorporated into planetesimals.

Comparisons of Run Set pairs \gen{HM} \& \gen{O}, \gen{P-I} \& \gen{P-IO}, and \gen{I} \& \gen{IO} show how the extended disk impacts the final realizations.  In general, the more extended disk leads to fewer planets forming inside of 1 au, though those planets are more massive than those formed in the more truncated disks (Fig. \ref{f:final}c).  This is likely due to the greater amount of mass in embryos and planets at larger distances, which can serve to excite the orbits of bodies close in, as well as scatter mass inward to be accreted by the planets growing in this region.  The larger excitations among the growing bodies lead to a greater spacing of the planets inside of 1 au, which -- in addition to reducing the number of such bodies -- leads to increases in the ratio of orbital periods.  Further, mutual inclinations increase as the orbits become more excited.

Thus, the amount of material in the outer disk affects the properties of planets formed inside of 1 au, consistent with the study of \citet{ballard16}.

\subsection{Effects of Surface Density Profile}\label{s:model:sdp}

The initial surface density profile of solids sets the total mass of embryos and planetesimals initially present in the simulation and how that mass varies with distance from the star.  As described above, this is usually defined as a power-law, and thus we examined two variations of how the surface density profile of solids impacted the final realization of planets.  Different surface densities may arise as disks have their masses distributed in different ways, or as radial migration of solids leads to the large-scale redistribution of solids prior to planetesimal formation.

In the first case, we considered how the surface density steepness, or exponent of the power-law, impacted the outcome of our models.  These effects can be seen by comparing the realizations of \gen{O}, \gen{O-p1}, and \gen{O-p2} where the surface densities profiles varied as $r^{-\frac{3}{2}}$, $r^{-1}$, and $r^{-2}$, respectively. The mass ratios of adjacent planets directly correspond to the slope of the surface density distribution. A slope of $r^{-2}$ leads to roughly equal-mass planets ($\mathcal{M} = 1$), while a shallow slope of $r^{-1}$ leads to the outer planets being more massive ($\mathcal{M} \approx 1.7$).

Comparing the realizations of these different runs, it is seen that more planets, with somewhat higher masses, are produced in the steeper surface density profiles, a result also seen by \citet{raymond05}. While the total amount of mass in the simulations is greater in the shallow surface density profiles, as they extend out to 4 au, the amount interior to 1 au, where the surface density is normalized to $50$ g/cm$^2$, is greater in the steep surface profiles, thus providing more raw materials for planets to form at smaller heliocentric distances.  The greater mass at larger heliocentric distances in the shallower profiles still impacts the dynamics of the interior planets, exciting them to higher inclinations and resulting in greater radial scattering of the interior planets, leaving the inner-most planet further from the star (Fig. \ref{f:final}d).

We also examined how the total mass of solids present in the simulations affected the properties of the final planetary system.  Model Runs \gen{O}, \gen{O-s22}, \gen{O-s10}, and \gen{O-s5} all had the same surface density steepness (varying as $r^{-\frac{3}{2}}$), but the surface density at 1 au varied, yielding different amounts of total mass to be accreted into the planets.  As shown in Table 2 and Figure \ref{f:final}, the higher surface density runs yielded more massive planets as more material was present to be accreted.  The more massive planets, due to their greater gravitational effects, cleared a larger amount of area around their orbits, leading to larger spacing between the planets, and thus fewer of them.  As the total mass decreased, the amount of dynamical excitation through the system decreased, with very weak gravitational interactions taking place, leaving a large number of low mass planets behind.  Similar outcomes were reported by \citet{kokubo06} and by \citet{raymond07} and  \citet{ciesla15} in looking at the formation of planetary systems around stars of different masses: greater numbers of smaller planets were produced around stars that had lower protoplanetary disk masses.  The trend of less gravitational excitement of bodies in the lower mass systems also leads to predictable results for the other quantities, with lower values for inclinations,  inner planet periods, and period ratios in these same systems.

\subsection{Effects of Gaseous Disk}\label{s:model:gas}
The simulations discussed thus far have been performed assuming that the gas from the protoplanetary disk was gone by the time this stage of planetary assembly began. However, as discussed above, some of this growth may have actually occurred  early in the evolution of the system while gas was still present.  Such early growth is necessary in our Solar System to explain the formation of the gas giants through core accretion \citep[e.g.][]{pollack96,hubickyj05} and the possible early formation of Mars relative to the typical lifetime of protoplanetary disks \citep{dauphas11}. In addition, low bulk densities of mini-Neptunes \citep{wu13,hadden14} and the presence of a valley in the planet radius distribution \citep{fulton17,owen17} indicate that the \textit{Kepler} planets likely formed in a gaseous environment.

Including disk gas in our simulations would have a dramatic effect on the dynamical evolution of planetary systems as gravitational interactions between embryos and the disk would lead to angular momentum exchange and the inward migration of the embryos. While a single migrating planet would stall at the inner edge of the gas disk, convergent migration of planets leads to inner planets in run sets migrating inside the disk inner edge, for example for run set \gen{M-s50} in Fig. \ref{f:final}f (see also \citealt{carrera19}).

In addition to migration, the gas also serves to damp down the eccentricities and inclinations of embryos.  This is reflected in the low mutual inclinations of planets in run sets \gen{M-s50}, \gen{M-s25}, and \gen{M-s10} (Fig. \ref{f:tablehist}.) and the large number of planets that form (as they do not get excited onto crossing orbits).  As a result, the masses of individual planets tend to be much lower, but there is a greater number of planets in a given system.  Further, the gas-driven migration and damping often drives planets into quasi-stable structures, thus such planets will settle into tightly-packed systems with low period ratios.

\begin{figure}
\includegraphics[width=\linewidth]{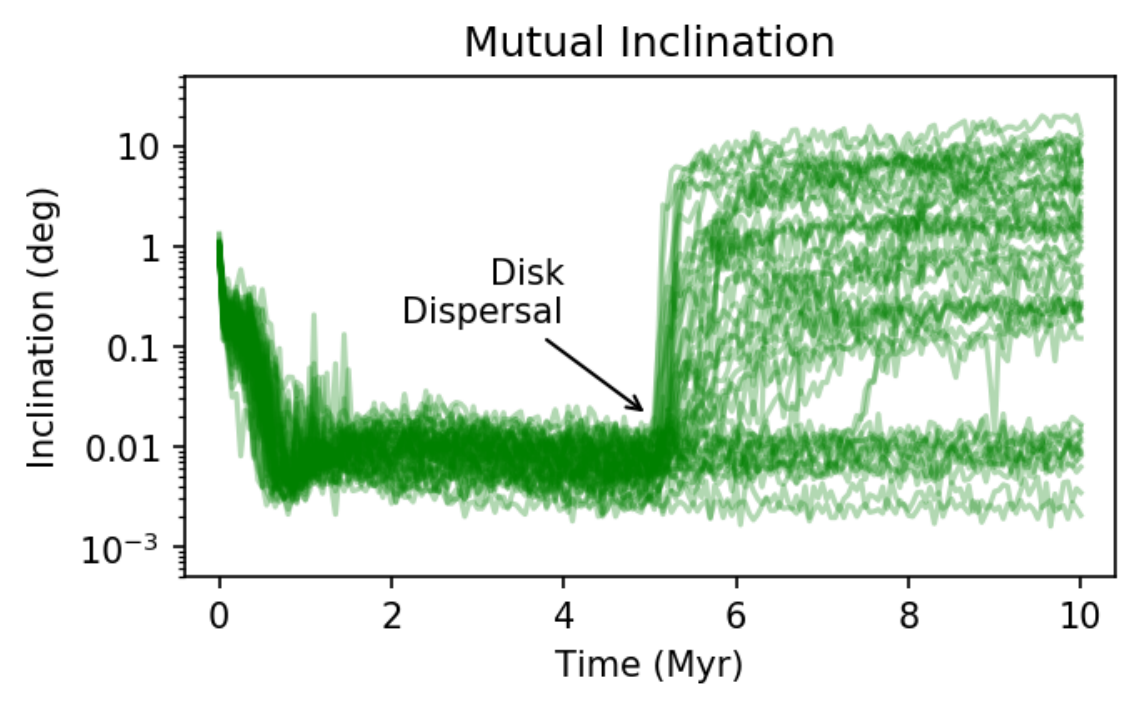}
\caption{Time-evolution of mutual inclinations in run set \gen{M-s22}, which includes gas damping and migration. The disk disperses at $5$ Myr, and system-wide instabilities occur within a few million years for a fraction of systems, after which systems remain fairly stable.
\label{f:evolution}}
\end{figure}

\begin{figure*}
\centering
\includegraphics[width=0.9\linewidth]{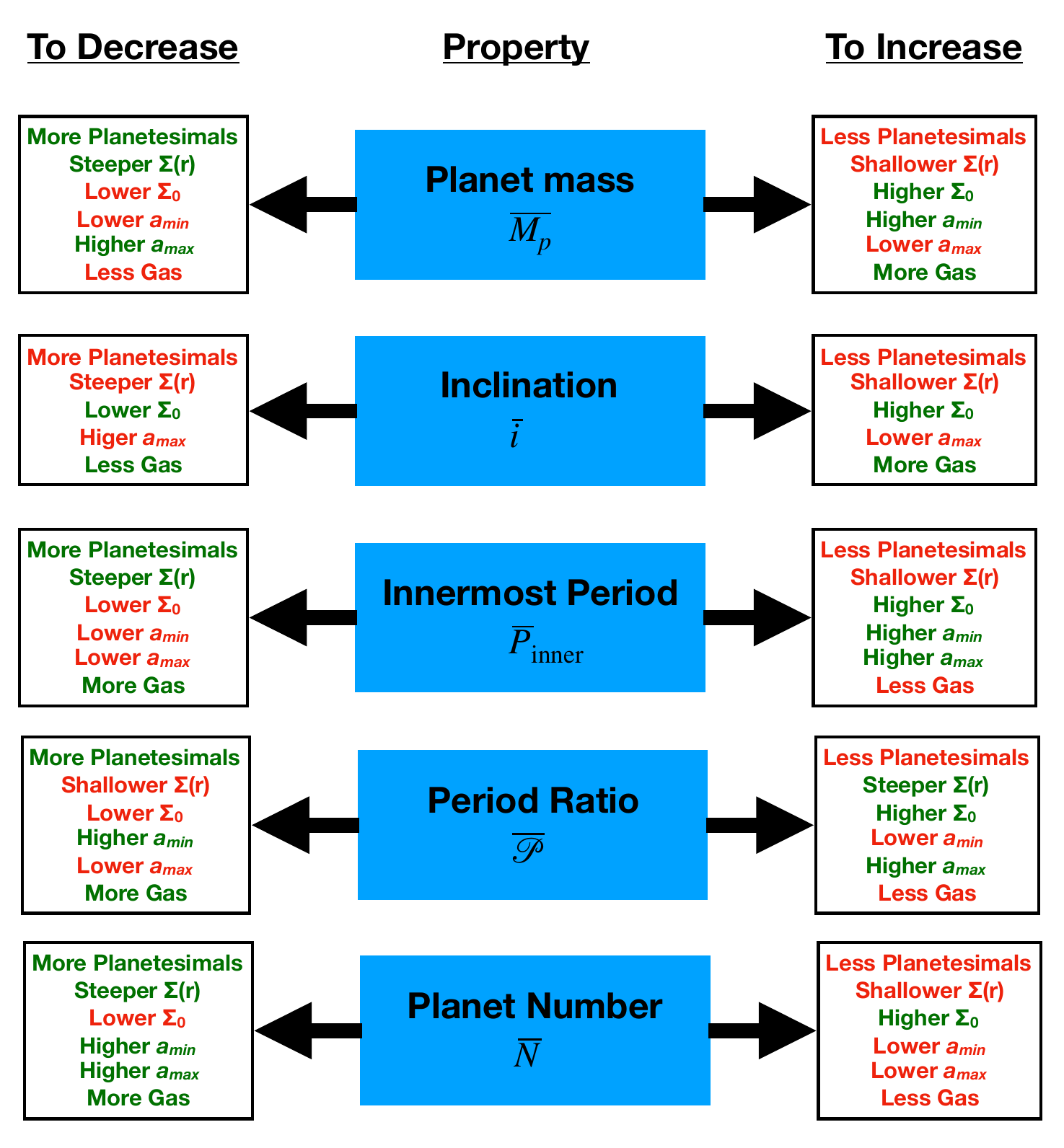}
\caption{Summary of property dependence on N-body initial condition.  
\label{f:nbody_summary}
}
\end{figure*}

While some level of accretion may occur in the presence of gas, protoplanetary disks are estimated to retain significant gas for only 3-5 Myr before the gas is lost via photoevaporation or other processes \citep[e.g.][]{ercolano17}.  Here we assume the gas is present for $\sim$5 Myr, though the gas density does decrease over time (before rapidly disappearing at the end).  In the absence of the gas, dynamical interactions and accretion occur as in the previously described cases. In some cases, these interactions are minimal, leaving a tightly-packed, stable system of planets.  In other cases, the dynamical interactions are such that the systems go through a period of instability, leading to large increases in mutual inclination.  Thus, in our simulations, as in \citet{izidoro17,izidoro19}, we generally see a dichotomy in this outcome (Fig. \ref{f:evolution}), with a bimodal distribution as  some systems have very low mutual inclinations while others have significantly elevated values, with few systems with mutual inclinations in between. While this is not readily apparent when looking at the median values in Table \ref{t:numbers}, it is more obvious when looking at the distributions of run sets \gen{M} shown near the bottom rows of Figure \ref{f:tablehist}, especially for \gen{M-s22}. Although most of our simulations are relatively short (~10 Myr), the result of the bimodal inclination distribution is likely to hold over the lifetimes of the systems: Running the simulations up to 100 Myr did not lead to a large number of these low-inclination systems becoming dynamically unstable and increasing their mutual inclinations. 

It is worth noting that the presence of gas can impact the relations that develop compared to those cases where gas is not present.  For example, in the higher mass surface density runs with no gas, we find that higher inclinations develop as a result of the gravitational interactions between the planets; those cases with lower surface densities have systems with lower inclinations.  The opposite is true when gas is present.  This is likely due to accretion occurring more rapidly when surface densities are high; given that a large amount of the gravitational interactions would take place early in the massive cases, the gas is still present to damp down any excitation that develop.  As models with lower surface densities would have more protracted planet growth, the gas dissipates before planetary assembly has progressed to a similar level as in a high surface density disk, allowing the planets in the system to become more inclined. Alternatively, more massive planets experience stronger tidal damping from the gas disk. 

\subsection{Summary of different effects}

The chaotic nature of planetary accretion means that even with similar starting conditions, the final properties of the emerging planetary systems may differ significantly from one another (Figure \ref{f:individual}).  However, given a large enough suite of simulations, it is clear that trends and relationships between the final properties and the initial conditions can be recognized (Figs. \ref{f:tablehist}, \ref{f:final}).  In this section, we have identified and explained how those trends develop for a set of final planetary system properties that can be inferred from the \emph{Kepler} planetary systems.

 Figure \ref{f:nbody_summary} summarizes how changes in individual parameters would lead to changes in the distribution of planetary systems that form.  While none of our specific \emph{Run Sets} exactly reproduce the properties of the \emph{Kepler} systems, it is possible to use the trends and relationships that we have outlined here to adjust the parameters and components to provide a better match between the modeled and real systems.  However, it is necessary to ensure that a proper comparison between the model realizations and \emph{Kepler} systems is done, requiring a rigorous consideration of observational biases.

\begin{figure*}
\includegraphics[width=\linewidth]{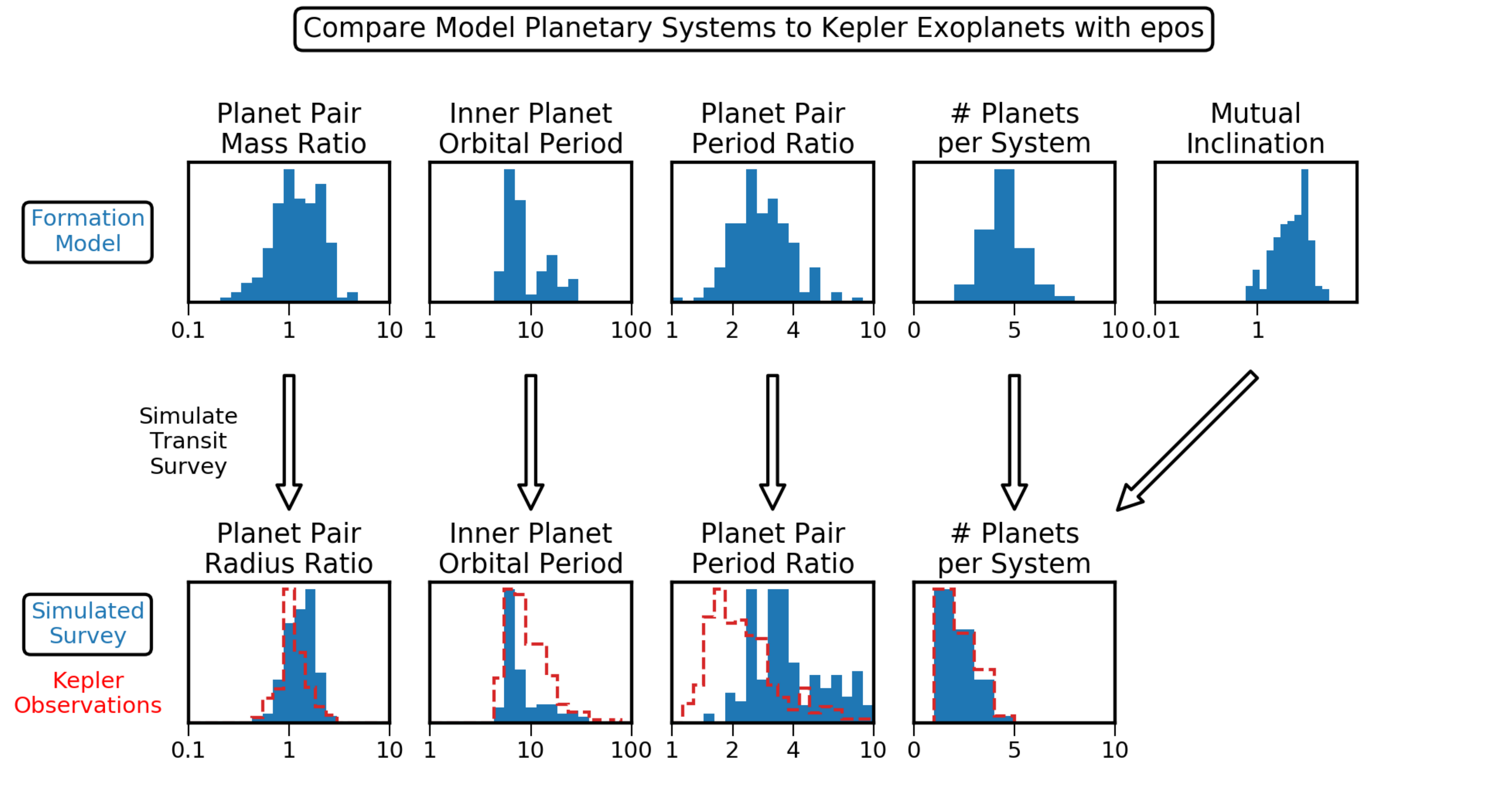}
\caption{Comparison of simulated planetary systems with observed \textit{Kepler} exoplanet systems. The top row shows the distribution of planetary system properties across run set \gen{HM}. The bottom row shows a simulated \textit{Kepler} survey of those systems with \texttt{epos} (blue), compared to the actual observations from \textit{Kepler} in red. Note that planet masses from the formation model are converted to planet radii using a probabilistic mass-radius distribution before applying detection biases. Planetary system mutual inclinations are not a direct observable and are mainly constrained from the observed number of planets per system. 
\label{f:arrows}}
\end{figure*}

\section{Statistical Comparison to the Kepler Planet Population}
In the previous section, we discussed how various accretion scenarios lead to significantly different distributions of planetary system properties.  In general, the model systems did reproduce some of the features of the \textit{Kepler} populations, yielding properties that overlapped with what is observed -- in particular for the model that includes planetesimals, \gen{P}. However, these comparisons must be made with care, as they are done with knowledge of the full properties of the planetary systems that formed, whereas our knowledge of the observed \textit{Kepler} system properties is limited by a number of observational biases. 
 Thus, a detailed comparison between the systems produced in our models must account for these limitations in order to determine which values of the parameters for our synthetic populations would actually be detected (Fig. \ref{f:arrows}) and whether the \textit{Kepler} observarions can be used to distinguish between the different accretion scenarios.

To do the forward modeling, we use \texttt{epos} version 3.0 \citep{eposv3}, the Exoplanet Population Observation Simulator, whose structure and detailed methodology have been described in \citet{epos1,epos2}. In short, \texttt{epos} performs a synthetic transit survey by assuming that the planetary populations produced by models make up the real distribution of planets, then performs a simulated transit survey of these systems, assuming that systems are randomly oriented along our sight lines and applies all detection biases of the \textit{Kepler} spacecraft.  We then compare the properties of the planets detected in the simulated survey to those of the systems detected by \textit{Kepler}, which are representative of planetary systems around solar-mass stars.  This approach, illustrated in Figure \ref{f:arrows}, allows for an apples-to-apples comparison between simulated planetary systems and the data, without having to parameterize the distribution of planetary system properties. 

We use four observational diagnostics of planetary system properties. The first three diagnostics were defined in \citet{epos2}, namely the observed frequency of multi-planet systems (\S \ref{s:multi}), the location of the innermost observed planet in each system (\S \ref{s:inner}), and the period ratio of adjacent planets (\S \ref{s:ratio}).
The fourth diagnostic is the size ratio of adjacent planets (\S \ref{s:cluster}), which has recently been shown to be strongly clustered in \textit{Kepler} planetary systems \citep{hsu19,sandford19}.

An example of this observational comparison for run set \gen{HM} is shown in Figure \ref{f:arrows}. The five panels in the top row show the distribution of planetary system properties from the planet formation models that are also shown in Fig. \ref{f:tablehist}. The bottom panel show the four observational diagnostics calculated by \texttt{epos} using a forward modeling approach of the detection biases, compared to the same diagnostic calculated for the exoplanet candidates from \textit{Kepler}.

\begin{figure}
\includegraphics[width=\linewidth]{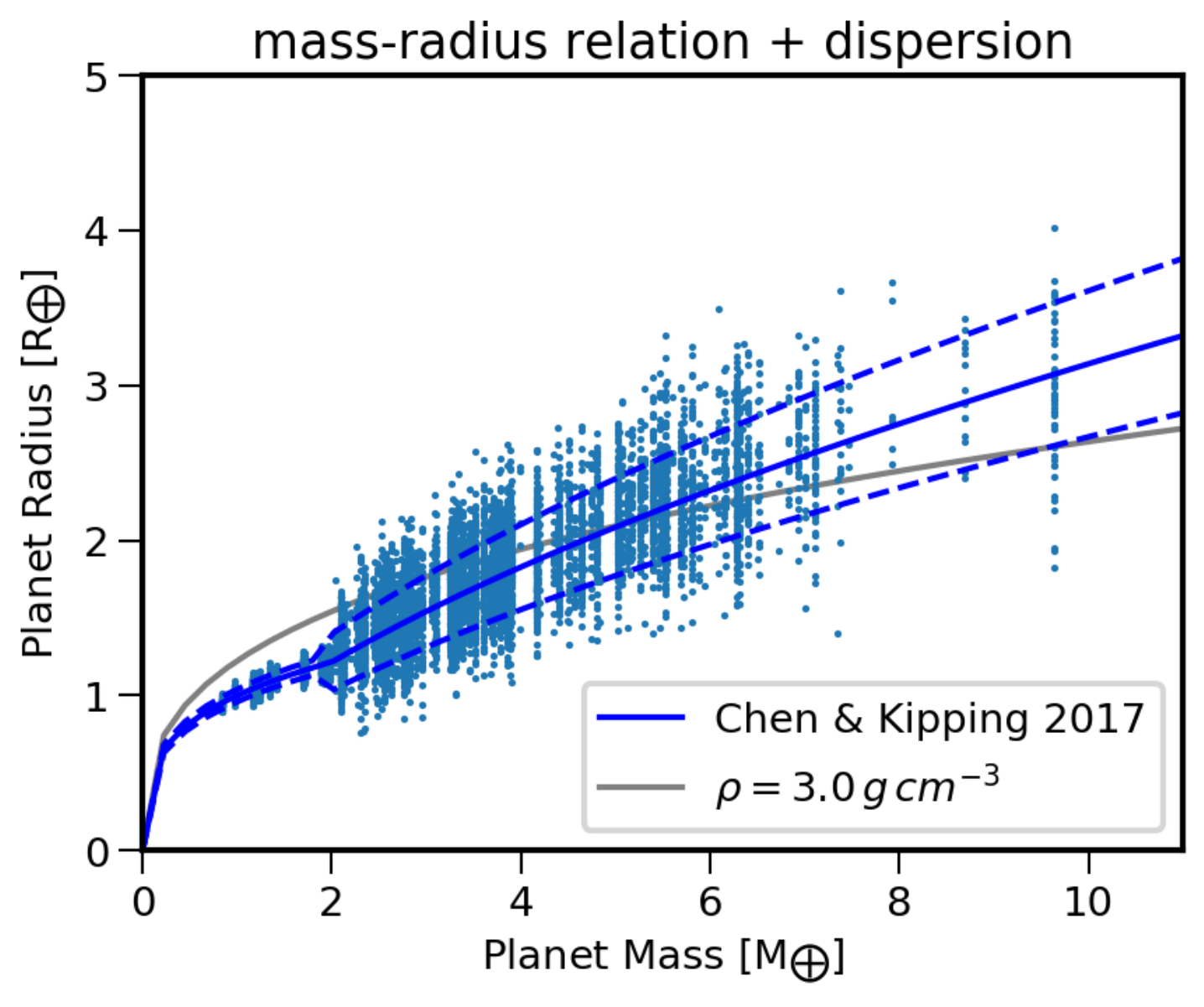}
\caption{Mass-Radius distribution of simulated planetary systems for \gen{HM} (blue-ish). The planet radii are drawn using a Monte Carlo simulation from the distribution of \citet{chen17} (dark blue), where the $1\sigma$ dispersion is indicated with the dashed lines. The constant density model assumed in the N-body integrator is shown in gray, which $\rho=3\,g/cm^3$.
\label{f:MR}
}
\end{figure}

An extra step that we perform compared to \citet{epos1,epos2} is that the planet radii are calculated from the simulated masses using the planet mass-radius relation from \cite{chen17}. Figure \ref{f:MR} shows this relation and its $1\sigma$ dispersion. For the full forward modeling in the synthetic survey, we draw the radii from a log-normal distribution centered on the best-fit mass-radius distribution using a Monte Carlo simulation. For reference, the constant density of 3 g/cm$^3$ as used in the N-body simulation is also shown, which appears to be a reasonable value for mini-Neptunes as well. 

For most runs, especially those where the inner disk edge $a_\text{min}$ is set to $0.05$ au, the subset that is compared to contains the majority of detected planet candidates, and thus a detailed statistical comparison can be made. 
In general, we find that the run sets with significantly different distributions of planetary system properties (Table \ref{t:ks}) are also distinguishable in these observational diagnostics. This demonstrates that while some information on planetary system architectures is inevitably lost due to transit survey biases, different formation scenarios can still be constrained from the observed \textit{Kepler} multi-planet systems.

For certain model runs, however, the overlap with the \textit{Kepler} planet candidates is limited because few planets are formed at short orbital periods and the probability of detection of multiple planets per system is extremely low. In those cases the sub-set of observed planetary systems is too small to allow a meaningful comparison on multi-planet statistics. This is the case for runs with an inner disk edge at $a_\text{min}=0.2$ au (\gen{I}, \gen{IO}, \gen{P-I}, \gen{P-IO}) and the run set with a very flat surface density power-law (\gen{O-p1}) that need to be compared to \emph{Kepler} planet candidates outside of a $\sim40$-day period. The data in this range contains only a handful of observed 2-planet systems, making them less likely to be relevant to the collection of planetary systems we are interested in.  Thus, we omit these simulations from the following analysis.

\begin{figure}
\includegraphics[width=\linewidth]{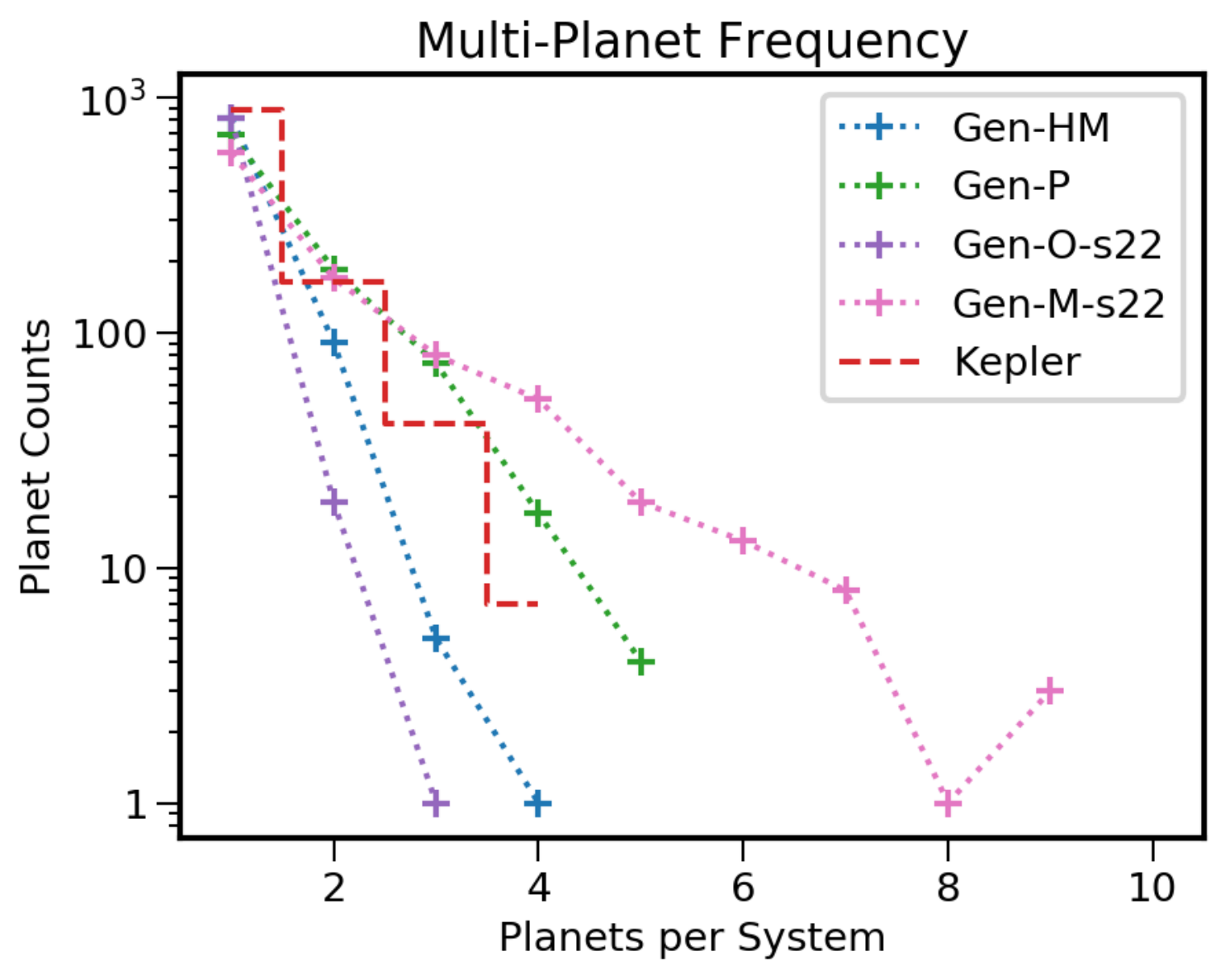}
\includegraphics[width=\linewidth]{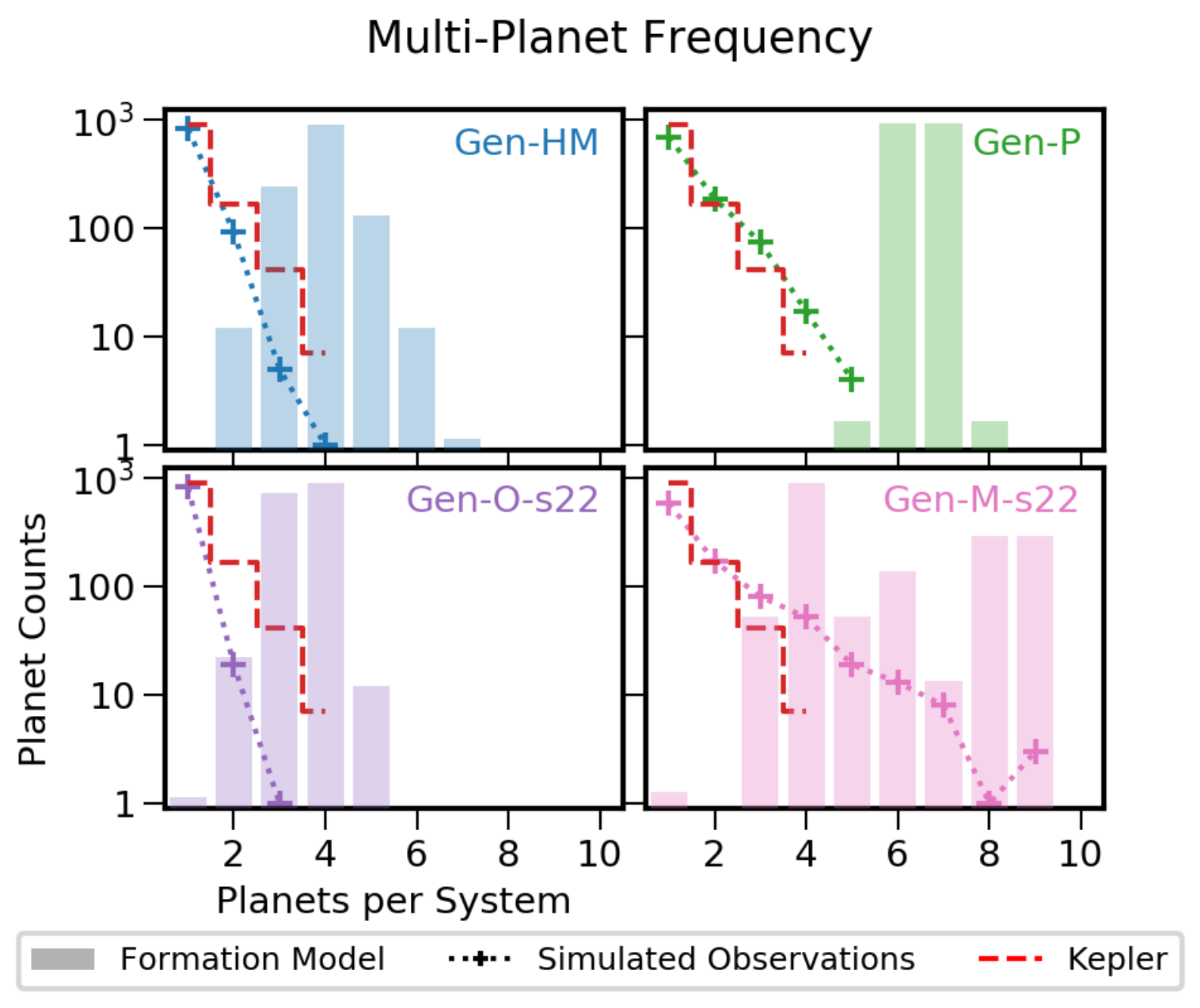}
\caption{Observable multi-planet frequencies for different accretion scenarios. The red dashed line shows the observed frequencies from \textit{Kepler} in the same range of planet radii ($1-3\,R_\oplus$) and orbital periods ($5-300$ days) as the simulated planets. 
The top panel shows the observable multi-planet frequencies for four different models to highlight the role of additional stirring when enlarging the disk outer edge (\gen{O-s22}) compared to the base simulation (\gen{HM}) and the role of additional damping by planetesimals (\gen{P}) and gas (\gen{M-s22}).
The bottom panel shows how the multiplicity of planetary systems in the formation model (bars) is completely different from the observable distribution of transiting systems (plus signs) due to non-transiting planets. 
\label{f:epos:multi}}
\end{figure}

\subsection{Number of Planets Observed in Systems}\label{s:multi}
The first diagnostic discussed here is the relative frequency of observed multi-planet systems, i.e., the observed number of stars with $m$ planets. 
Because most planets in multi-planet systems do not transit, this diagnostic does not always reflect the true multiplicity of systems, $N$. 
While most exoplanet host stars in the \textit{Kepler} survey only have a single transiting planet detected, statistical studies show that most of these are members of planetary systems with multiple non-transiting planets \citep[e.g.][]{lissauer11,zink19}.
Instead, the relative frequencies are most sensitive to the mutual inclinations, with a decreased chance of observing more planets per star if the mutual inclinations are higher. 

To show what the observable signature of the different planet accretion scenarios would look like, we compare four run sets that all produce planets in the same size range ($1-3\,R_\oplus$) and orbital period range ($5-300$ days) in Figure \ref{f:epos:multi}. Run \gen{HM} represents an embryo-only scenario, \gen{P} a planetesimal scenario, \gen{O-s22} a large disk scenario, and \gen{M-s22} a migration/damping scenario. 
The simulated distributions of multi-planet frequencies of these four run sets are significantly different at high confidence ($p<0.01$), showing that these diagnostics can indeed be used to discriminate between the different accretion scenarios shown in Figure \ref{f:tablehist}.
The observational signature of these runs are then compared to the observed frequencies of multi-planet systems from \textit{Kepler} in the same ranges of planet radii and orbital periods.

The embryo-only simulations of \gen{HM} under-predict the frequencies of observed 2, 3, and 4-planet systems compared to \textit{Kepler}, likely because their mutual inclinations are significantly larger than the 2 degrees typically inferred for \textit{Kepler} multi-planet systems (see also Fig. \ref{f:tablehist}). While \cite{hansenmurray13}, on which the \gen{HM} runs are based, found a good match to the observed ratio of multi-planet systems, they compared their simulated planetary systems to an earlier version of the \textit{Kepler} catalog that had fewer 3- and 4-planet systems. Thus, our model results are consistent with theirs, even though we find no good match to the updated \textit{Kepler} systems. 

Models from the \gen{O} series in which the outer disk is extended to a semi-major axes of $4$ au, of which \gen{O-s22} is shown here, under-predict the number of observable 2 and 3-planet systems compared to the observations by a larger amount than \gen{HM}. This is likely due to the increased dynamical stirring from the outer disk mass leading to larger mutual inclinations, which decreases the chance of multiple planets transiting  from a given viewing angle (see also \citealt{moriarty16}). These conclusions extend to models \gen{O}, \gen{O-p1}, and \gen{O-s10} (not shown) which have a similar distribution of planet frequencies and are not distinguishable in this diagnostic. The multi-planet frequencies of model \gen{O-p2} are more similar to those of \gen{HM} because less mass is added to the outer disk, minimizing its impact on the mutual inclinations and hence the multi-planet frequencies. The model with the lowest surface density (\gen{O-s5}) has less dynamical stirring and hence lower mutual inclinations, but a direct comparison with the data is difficult because the small planet sizes yield a very small sample of detected multi-planet systems to compare to.

A better match to the observed multi-planet frequencies is seen in simulations that include planetesimals where dynamical friction reduces the mutual inclinations. \gen{P} has a larger number of predicted 2--5 planet systems compared to \gen{HM}, and over-predicts the observed frequencies by a small fraction. The ratio of 3/2, 4/3 and 5/4 planet systems is the same as observed, and hence this model would closely match the \textit{Kepler} data if a population of single or highly inclined planets would be included to account for the \textit{Kepler} dichotomy, as described in \citet{johansen12}, \citet{hansenmurray13}, and \citet{ballard16}.

The inclusion of damping by gas leads to mutual inclinations that are so low that the runs over-predict the frequency of multi-planet systems. For example simulation \gen{M-s22}, shown in Figure \ref{f:epos:multi}, which has planets in a similar size range as \gen{HM} and \gen{P}, over-predicts the number of 3- and 4-planet systems. Moreover, it predicts a large number of 5-8 planet systems be observed which are not present in the \textit{Kepler} data in this size and period range.  Simulation run sets that include migration in combination with higher and lower surface densities (\gen{M-s50} and \gen{M-s10}, respectively) similarly over-predict the number of observable multi-planet systems due to their low mutual inclinations.

\subsection{Period Ratios of Adjacent Planets}\label{s:ratio}

\begin{figure}
\includegraphics[width=\linewidth]{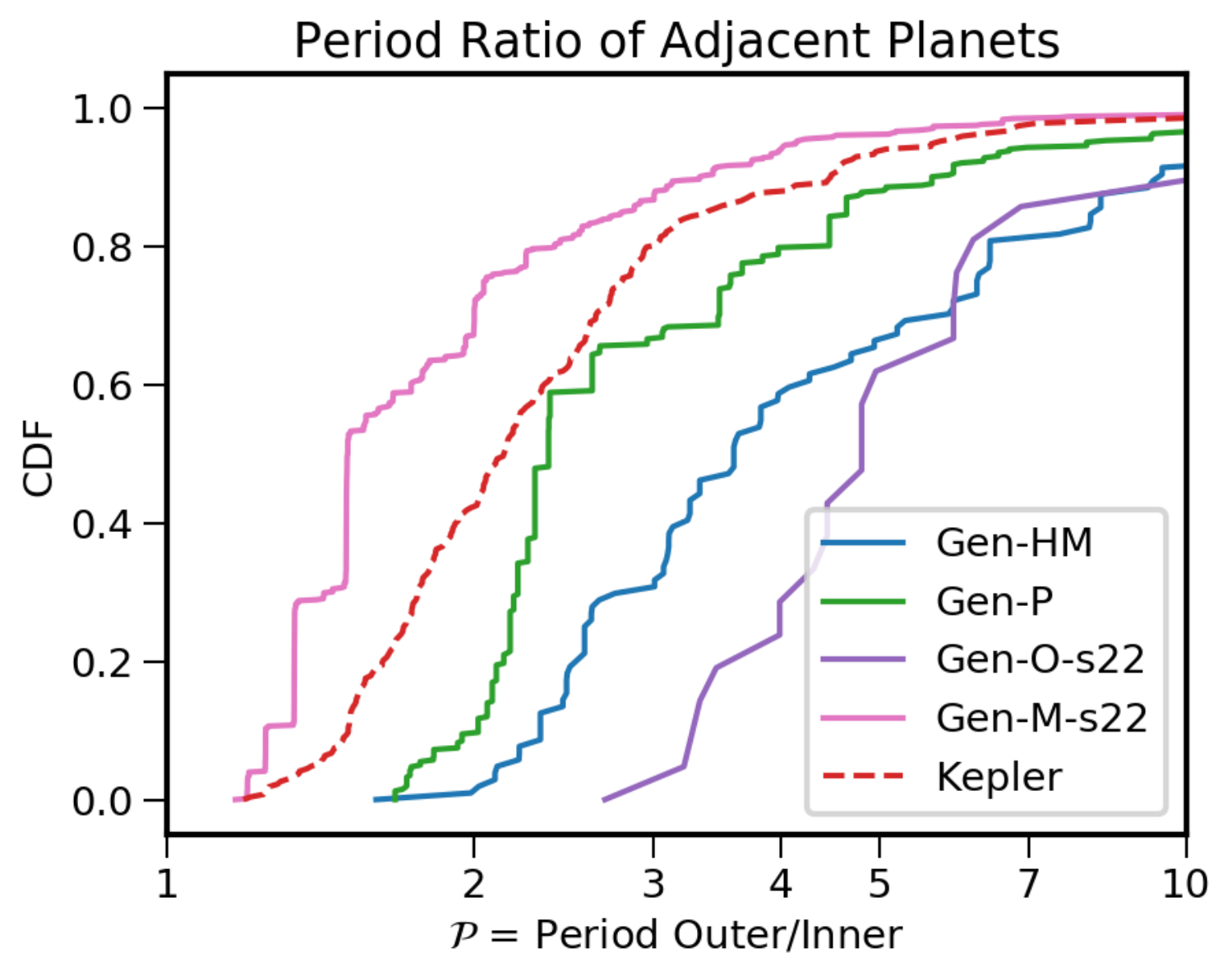}
\includegraphics[width=\linewidth]{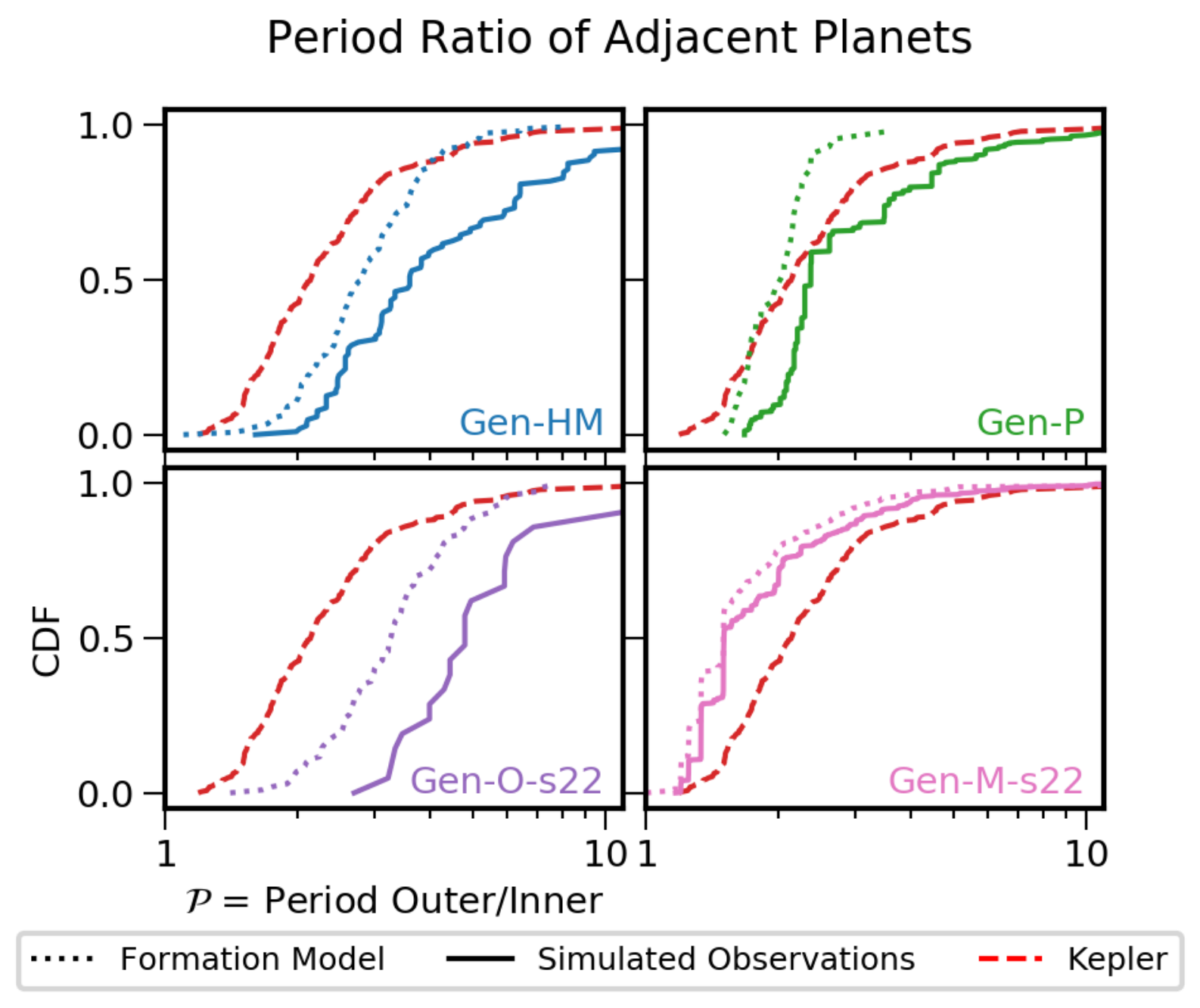}
\caption{Period ratio distributions in a simulated survey of the different models (solid lines) compared to the  \textit{Kepler} observations (dashed line). 
These comparisons highlight the role of additional stirring when enlarging the disk outer edge (\gen{O-s22}) compared to the base simulation (\gen{HM}) and the role of additional damping by planetesimals (\gen{P}) and gas (\gen{M-s22}).
\label{f:epos:dP}}
\end{figure}

The second diagnostic discussed here is the period ratio distribution of detected planet pairs, which mainly traces the orbital spacing between planets. This period ratio is defined as $\mathcal{P}=P_{k+1}/P_k$ where $k$ enumerates observed planets from inside out. This diagnostic is calculated for planets in the range of $1-3 R_\oplus$ and $5-300$ days. The detected period ratios of four models representative of different accretion scenarios are shown in Figure \ref{f:epos:dP} as a cumulative distribution to highlight the excess (or absence) of resonant features. The bottom panel shows how the detected distributions are biased with respect to the distribution predicted by the formation models. 
The observed distribution from \textit{Kepler} spans a range of period ratios from $\sim$1.3 to almost 10, with a median value at $\mathcal{P} \approx 2.2$. 

The period ratio distribution of the different accretion scenarios (Fig.~\ref{f:epos:dP}) follow a similar trend to that of the multi-planet frequencies: Simulations with the strongest damping (by gas) underpredict the observed period ratios and simulations with stronger dynamical interactions from larger disks progressively over-predict the observed period ratios.

The baseline simulation of \gen{HM} (and also \gen{O-p2}) has a median period ratio of $\sim3.3$ and the distribution is located at longer period ratios compared to the observations. The amount of dynamical stirring in the embryo-only simulations leads to planets that are spaced too far apart to be consistent with the \textit{Kepler} systems, as also noted by \cite{hansenmurray13}. The simulations with increased mass in the outer disk (\gen{O-s22}, \gen{O}, \gen{O-p1}) all have similar distributions with even larger period ratios due to the increased dynamical interactions. The simulations with a lower overall surface density (\gen{O-s10} and \gen{O-s5}) have smaller period ratios, but because these simulations also produce smaller planets, there are not enough observed planets to allow a statistically meaningful comparison with the \textit{Kepler} exoplanet population.

When dynamical friction from planetesimals are included in the simulations, the emerging planets are more tightly packed. For example the predicted period ratio distribution of simulation \gen{P} has a mean period ratio of $\mathcal{P} \approx 2.4$, just higher than the observed distribution from \textit{Kepler}. The shape of the distribution is roughly similar to the observed period ratio distribution, and is the closest match to the data of all the simulated systems. We did not explore if a larger number or mass fraction of planetesimals could further reduce the period ratios and provide an even better match to the observed systems.

Simulations including orbital damping by gas and planet migration, on the other hand, result in observable planetary systems that are too compact. For example, simulation \gen{M-s22} has a mean period ratio of $\mathcal{P}=1.5$ . 
The distributions also show strong features at the main orbital resonances (2:1 and 3:2) that are not present in the \textit{Kepler} data. Dynamical instabilities that break these resonances and widen the period ratio distribution are needed to provide a better match to the observed systems,  similar to recent results by \cite{izidoro19}. 

The solid surface density has a big impact on the observed period ratio distribution of planets formed in a gaseous disk. Simulations with a higher (lower) surface density of solids grow and migrate faster (slower) and therefore experience more (less) damping by gas and less (more) breaking of resonant chains. For example, simulation \gen{M-s50} with the highest solid surface density show stronger resonant features and smaller orbital periods and is therefore less consistent with the observed period ratio distribution. On the corollary, the simulation with the lowest surface density, \gen{M-s10}, has fewer planets in orbital resonances and a period ratio distribution that is only slightly shortwards of the \textit{Kepler} distribution. 
While the period ratio distributions of the three migration run sets are statistically distinguishable, they are all inconsistent with the observed distribution from \textit{Kepler}.

\begin{figure}
\includegraphics[width=\linewidth]{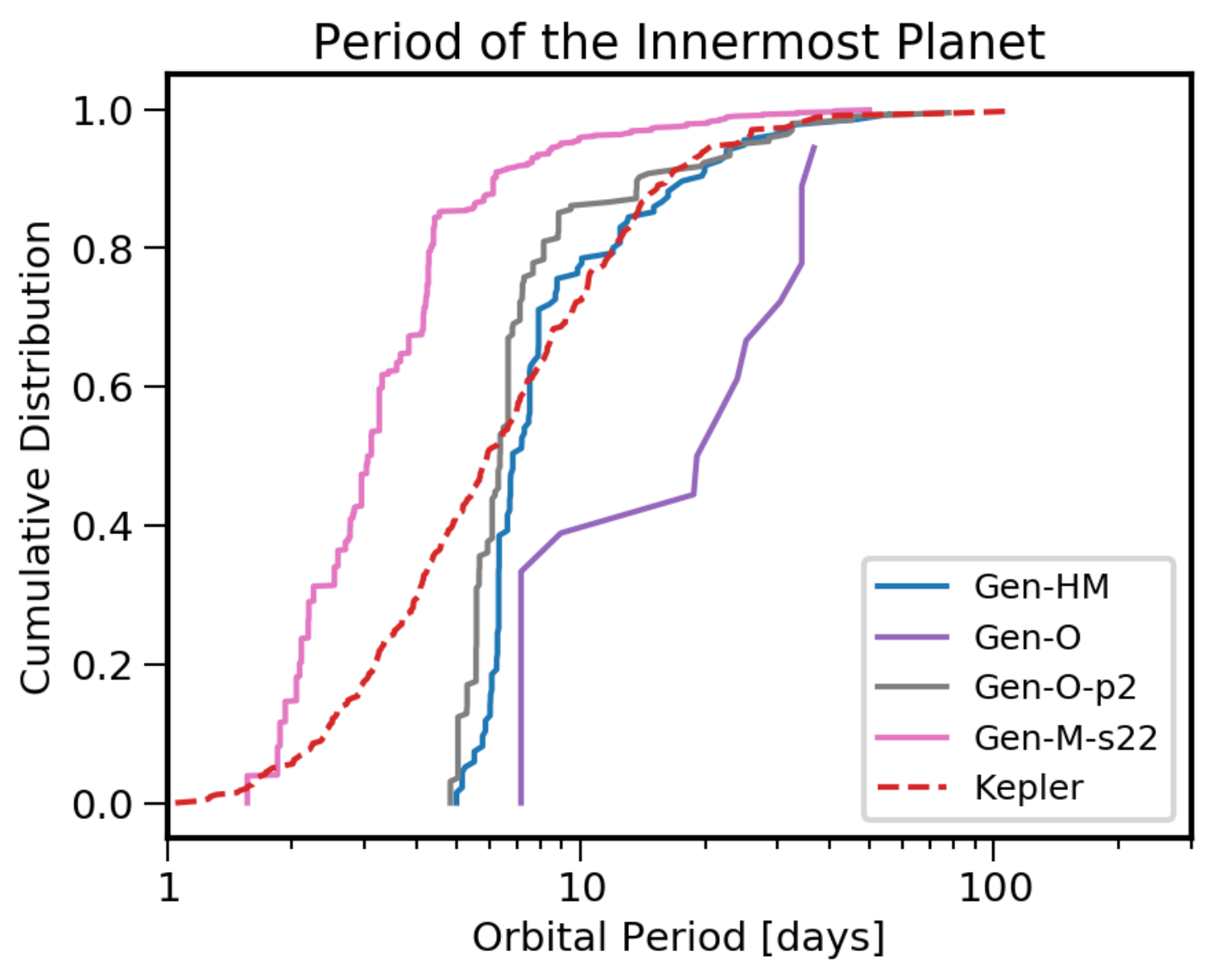}
\includegraphics[width=\linewidth]{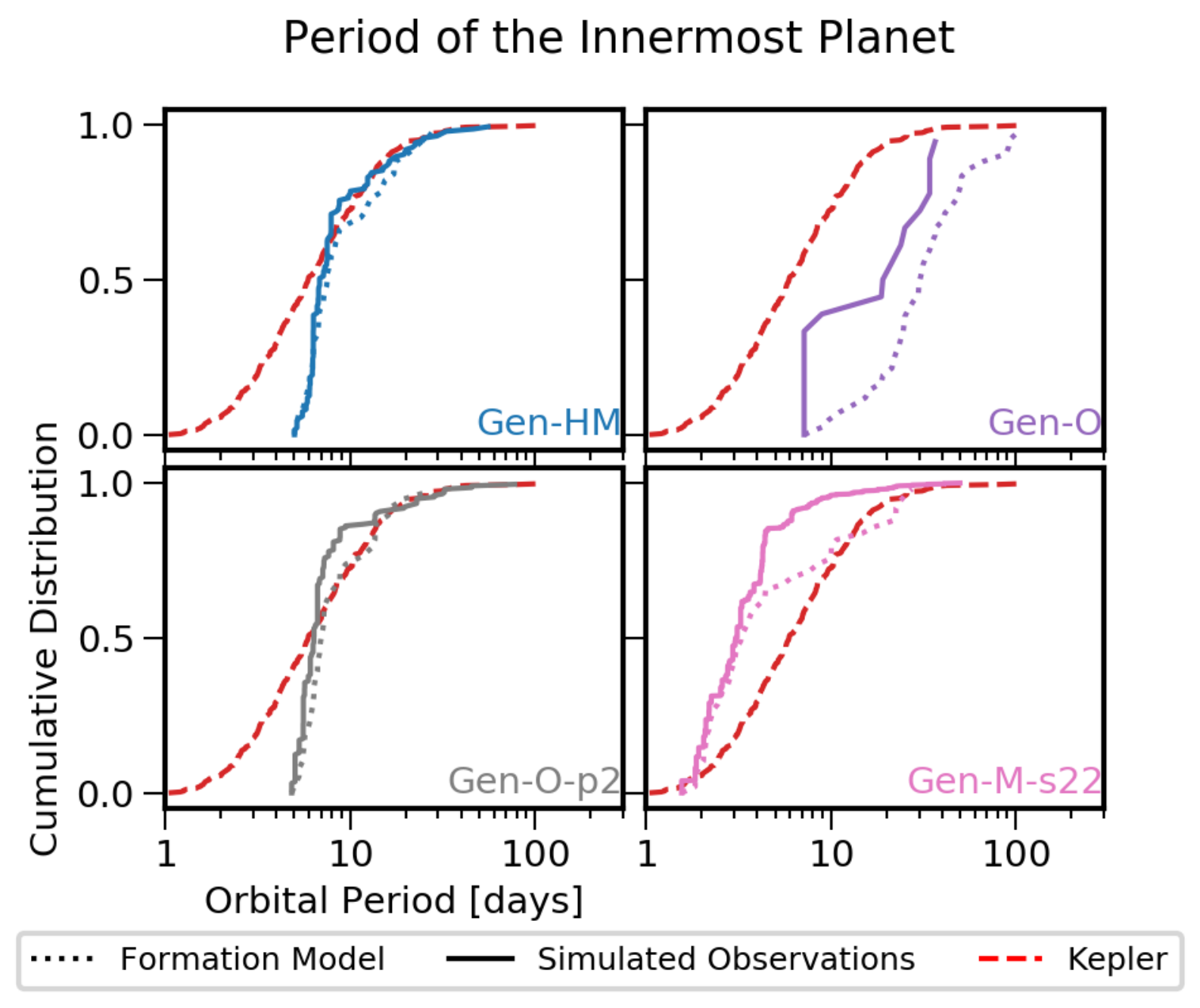}
\caption{Period of the innermost planet in an observable multi-planet system (solid). The observed distribution from \textit{Kepler} for planets in the same radius range ($1-3~R_\oplus$) is shown in dashed red.
These comparisons highlight the role that migration and disk size play in determining the location of the innermost planets. All models utilize the same inner disk edge of 0.05 au. % P=..
Planet migration moves the innermost planet inwards to the edge of the gas disk which is located at 0.05 au. A large outer disk results in inner planets at larger separations (\gen{O}) compared to a $1$ au disk (\gen{HM}) unless the outer disk contains less mass (\gen{O-p2}).
\label{f:epos:Pinner}}
\end{figure}

\subsection{Period of the Innermost Planet in Each System}\label{s:inner}
The third diagnostic discussed here is the location of the innermost detected planet in an observed multi-planet system. For each system with more than one planet detected in the simulated survey, we recorded the location of the innermost detected planet. The distribution of these values is then compared to the locations of the innermost planets in the \textit{Kepler} multi-planet systems.
Figure \ref{f:epos:Pinner} shows the inner planet locations for four different models in the range $1-3~R_\oplus$.
Note that this is a different selection of models than in Figures \ref{f:epos:multi} and \ref{f:epos:dP}, and was chosen to highlight the parameters that most affect the location of the inner planets.

The distributions before and after applying detection biases, shown in the bottom panel of Fig. \ref{f:epos:Pinner}, differ primarily because two detection biases are at play. First, inner planets at shorter orbital periods have an increased transit probability, which shifts the distribution of detected inner planets towards shorter periods compared to the instrinsic distribution of inner planets. Second, in a fraction of systems the inner planet is not transiting, and the innermost detected planet is located farther out, effectively broadening the distribution. \texttt{epos} takes both these biases into account when calculating the detectable distributions.

We find that the period of the observed innermost planet varies significantly even between simulations that have the same inner disk edge but differ in the other initial conditions (Fig. \ref{f:epos:Pinner}). The fiducial simulations of \gen{HM} and \gen{P} predict an observable inner period ratio distribution with roughly the same mean value as found in the \textit{Kepler} data, and are not statistically distinct from each other. However, in both cases the distributions are much narrower than observed.  This difference in the widths of the simulated and observed distributions could indicate that a range of inner disk locations, rather than a single value, is needed for this model to match the observed inner period distribution.
\cite{lee17} have suggested that the observed spread in stellar rotation rates of young stars leads to such a range if the inner disk edge is located at the radius corresponding to the stellar co-rotation.  This was also suggested by \cite{mulders15a} based on the stellar-mass (in)dependence of the period-break in the planet occurrence rates.

Increasing the mass available by extending the outer disk (models \gen{O}, \gen{O-p1}) increases the amount of dynamical interactions between planets in the inner disk and also pushes the location of the innermost planets outward (see Table \ref{t:numbers}). 
The latter is reflected in the simulated observables, where the median inner period of model \gen{O} is almost twice that of model \gen{HM} (Fig. \ref{f:epos:Pinner}).
The inner planet orbital period distribution is also wider than that of the fiducial model, indicating that a better fit to the data might be obtained with a larger disk but with a smaller inner disk edge. Simulation \gen{O-p2} also has an enlarged outer disk but less mass is added in this case because of the steeper surface density power-law, and this model therefore produces a comparable inner planet orbital period location distribution as run set \gen{HM}.

Adding gas to the disk makes planets migrate inward, pushing the observable distribution of innermost planets inward. The median inner period of model \gen{M-s22} is more than half that of the baseline model, with a median observable period at 3 days. The simulated distribution is narrower and located at shorter orbital periods as the observed inner planet distribution. A model where the inner edge of the gas disk is located further out at an orbital period of 10 days \citep[e.g.][]{epos1,carrera19}, and has a dispersion in locations, would likely provide a better match to the observed \textit{Kepler} systems.

\subsection{Radius Ratios of Adjacent Planets}\label{s:cluster}

\begin{figure}
\includegraphics[width=\linewidth]{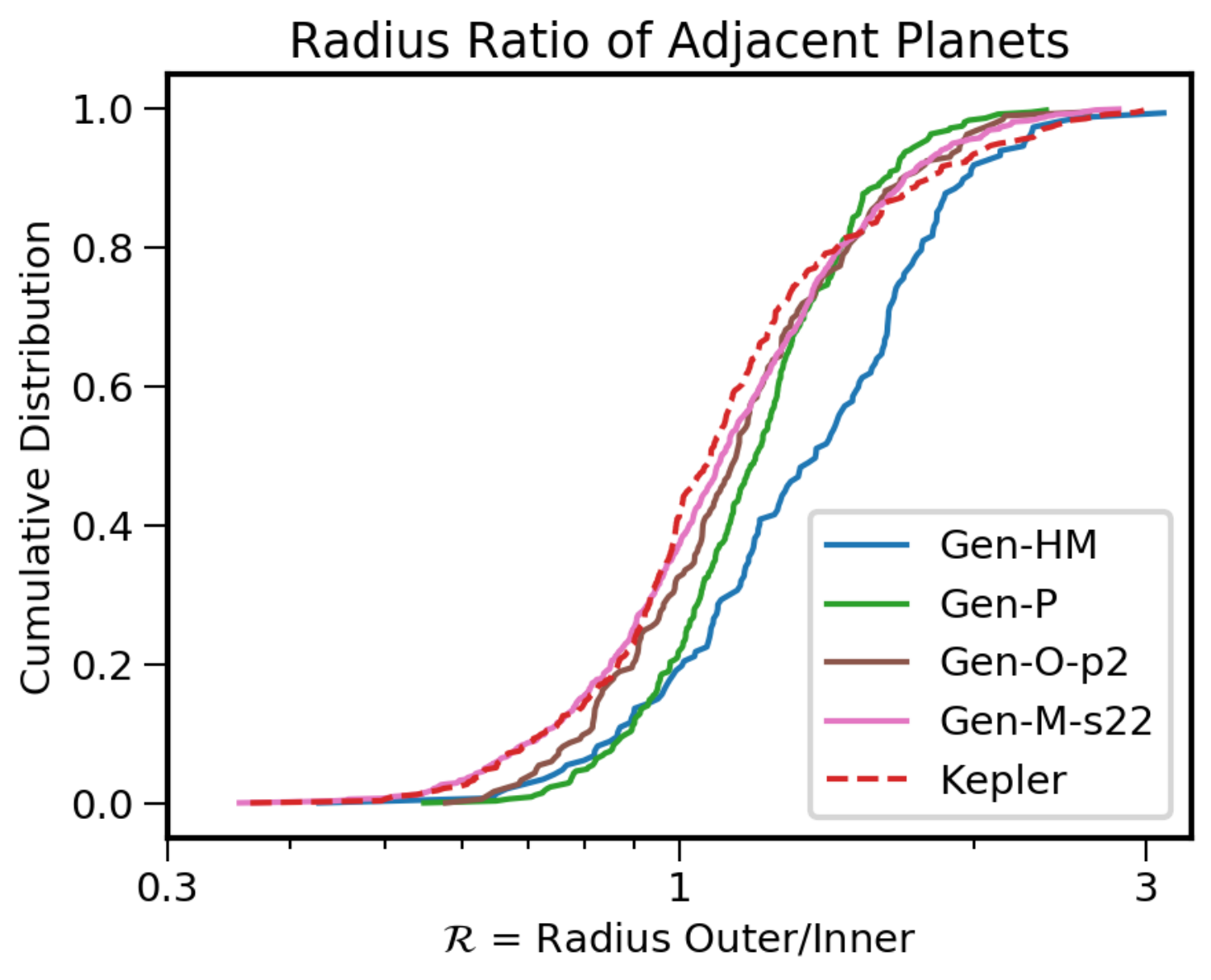}
\includegraphics[width=\linewidth]{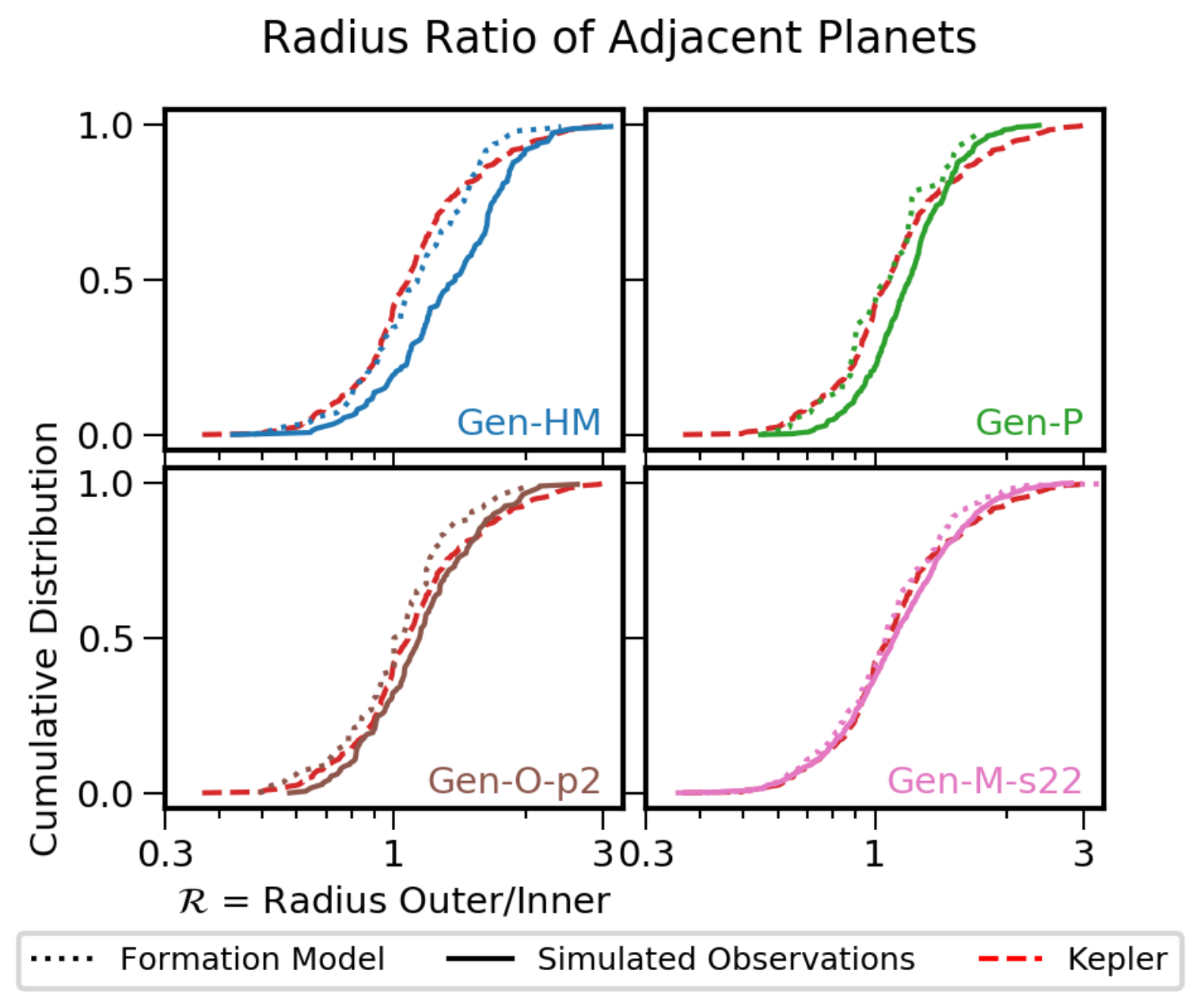}
\caption{Radius ratio distributions in a simulated survey of the different models (solid lines) compared to the  \textit{Kepler} observations (dashed line). 
\label{f:epos:dR}}
\end{figure}

The fourth diagnostic discussed here is the radius ratio distribution of adjacent detected planets, which mainly traces the clustering in planet sizes. This radius ratio is defined as $\mathcal{R}=R_{k+1}/R_k$ where $k$ enumerates observed planets from inside out. The top panel of Figure \ref{f:epos:dR} shows the cumulative distribution of radius ratios for four different models in the simulated survey. 
Note that this is again a different selection of models than in Figures \ref{f:epos:multi}, \ref{f:epos:dP} and \ref{f:epos:Pinner}, because the model parameters that affect the distribution of planet radii most are different. 
The observed distribution from \textit{Kepler} for planets in the range of $1-3\,R_\oplus$ and $5-300$ days spans a range of radius ratios from $\sim$0.5 to almost 3. The distribution is slightly asymmetric with a median at $\mathcal{R} \approx 1.1$. 

The bottom panel in (Fig. \ref{f:epos:dR}) shows how the size ratio distributions are affected by observation biases. The main detection bias here arises because the minimum detectable planet size increases with orbital period because fewer transits occur at large periods. A planet pair where the outer planet is larger is more likely to be detected than a planet pair where the outer planet is smaller. Thus, the size ratio distribution in the synthetic survey is shifted to larger size ratios than those in the planet formation model.

The radius ratios of the different accretion scenarios, subjected to observational biases, are shown in Figure \ref{f:epos:dR}. The simulated distributions typically have a similar shape as the observations in each scenario, indicating that the radius clustering of planet sizes observed with \textit{Kepler} is naturally produced by these simulations. But they differ in how much they are offset toward the larger planet radius ratios. In particular, simulations \gen{HM} and \gen{P} are offset to larger size ratios compared to the observations.

In a scenario without migration, trends in the size ratio of planets reflect the mass distribution in the initial disk. Assuming the width of a planet's feeding zone is proportional to its semi-major axis $a$, the surface area of the annulus that a planet accretes from scales as $a^2$. Models with a radial surface density exponent of $p=1.5$ are therefore expected to form planets whose mass increases with semi-major axis as $a^{0.5}$, and are thus expected to have size ratios that are on average larger than $1$. Planets formed in a disk with a radial surface density exponent of $p=2$ are expected to be of similar size at different distances from the stars.
We can indeed see that simulation \gen{O-p2} does not show a significant offset compared to the observations and is a better match than simulations \gen{HM} and \gen{P}.

Finally, we find that simulations with migration are almost a perfect fit to the observations, showing both the same offset and dispersion in the simulated planet radius distribution as in the observations. This is perhaps slightly surprising as the underlying surface density profile used in these simulations is the same as in the gas-free simulations. Likely, the inward migration and resonant trapping changes the spacing between forming planets such that the final planetary systems end up closer to equal in mass.

\section{Summary}
We have explored the effects that different initial conditions and model configurations have on the final stages of planetary accretion.  While the process itself is chaotic, we have identified relationships between the initial conditions and model configurations and the final planetary system properties, including the number of planets, their relative sizes, their orbital spacings, and their mutual inclinations that provide an imprint of their formation histories. Such statistical relationships can be used to inform us about the possible formation pathways of known exoplanetary systems.  Doing so, however, requires a detailed assesment of the observational biases to properly reveal the properties of these systems.

In applying our knowledge to the known \emph{Kepler} systems, we find that:
\begin{enumerate}
    \item Planets within planetary systems are typically similar in size, despite the chaotic nature of the late stages of planetary assembly. In every scenario we explore, whether that includes migration of planets in a gas disk or not, the distribution of planet masses within a system are consistent with the clustering of planet radii observed with \textit{Kepler}. 

    \item Planet formation did not occur purely through the assembly of planetary embryos as such systems would be too dynamically excited 
    compared to the real systems.  Leftover planetesimals that were present during the final stages of accretion would serve as a source of dynamical friction that provide better matches to the multi-planet frequencies and inclinations, though the simulations performed here still yield spacings between detected planets that were larger than observed.

    \item Planetary systems where a significant amount of accretion occurs within a gaseous protoplanetary disk, are too dynamically cold to match the observed exoplanet population, with period ratios that are too small and have an overabundance of multi-planet systems. In contrast, systems that undergo dynamical instabilities after the disk has dispersed are dynamically hotter than those that stay stable, and the planets' mutual inclinations within a system are closer to what is needed to match observed systems. It is, however, difficult to determine a priori which systems will go unstable and which will not.    
    
    \item The structure of the outer disk, outside of the region where planets are detected with \textit{Kepler}, influences the properties of planets and planetary systems formed closer to the star, indicating that planet formation is not always a strictly local process. Planets formed in more extended disks are more massive, located farther from their host stars and are less likely to be observed as multi-planet systems. 

\end{enumerate}

None of the simulations performed here simultaneously matched all four observational diagnostics of planetary systems, which are the relative frequencies of multi-planet systems, the orbital period ratios of planet pairs, the size ratio of planet pairs, and the locations of the innermost planet. However, the above trends between initial conditions and observables allow us to better estimate the conditions under which these systems formed. 

In regards to developing a collection of systems whose innermost planets are similar to those seen in the \emph{Kepler} systems, it is necessary to consider a range of inner disk locations. The embryo-only simulations are consistent with an inner disk edge centered at 0.05 au, but future simulations should consider implementing a range for the inner disk edge rather than a single value.
In more extended disks a single inner disk edge location provides a broader detected distribution that is more consistent with the data, though this inner edge has to be located closer in than 0.05 au. In simulations including migration and orbital damping by gas, the inner disk edge needs to be moved outward to $\sim0.1$ au to result in planetary systems consistent with what is seen with \textit{Kepler}.

Dynamical damping by gas in the cases studied here was too efficient, leading to too many observable multi-planets systems and planet orbital period ratios that are too low.  Similar findings were found in other studies: population synthesis models with a similar number of interacting seed cores (20-50) also find mutual inclinations of planetary systems that are too low \citep{epos2}; and pebble-assisted formation and migration models find that not enough systems go unstable after disk dispersal to increase their mutual inclinations to the point of reducing the number of observed planets and increasing the observed period ratios \citep{izidoro17,izidoro19}. Less efficient damping by gas may thus be necessary. For example, \cite{dawson16} and \cite{MacDonald2020} show that if planetary system assembly is delayed and happens when the disk gas is depleted by a factor 100 compared to the the initial gas mass, the amount of dynamical excitation matches closer to the observed systems. Although the underlying model we use has a gas density that decreases with time, we do not see this effect naturally arising from our simulations. We do see, however, that in disks that form smaller planets a closer match to the observed system properties is found. This is likely a result of these planets growing slower and, therefore, assembling when the gas density is lower or already dissipated, which, in turn, provides weaker or no orbital damping.  This could suggest that the final stages of planetary accretion generally took place relatively late in terms of the removal of disk gas. 

The general formation conditions outlined here allow us to make predictions about the formation pathways of habitable zone planets, and their potential chemical composition.   
For example, the likely presence of planetesimals during planetary accretion
would provide a means for materials from more distant regions to be delivered to planets that form closer to the Sun.  Such a process would be a natural means of delivering water and other volatiles to planets in the Habitable Zone.  

The presence of gas might limit such radial mixing of planetesimals, but the gas would also induce type I migration, 
allowing planetary embryos from the outer disk to drift inwards toward the Habitable Zone.  These embryos would also have formed in volatile-rich regions, allowing planets in the habitable zone to contain more water, carbon, or other bio-critical elements than if they were to have formed in situ. Since the amount of orbital damping by gas appears to be limited based on our simulation results, the influence of migration on planet composition is likely also reduced.

Quantifying the impact of different planet formation pathways on planet compositions requires an understanding of both the chemical composition in the disk as well as a model for how the distribution of material at different distances is assembled into planets. As such, we are making all simulated planetary systems discussed in this paper -- complete with their accretion history -- available in the \texttt{Genesis} database at \url{http://eos-nexus.org/genesis-database/}. The \texttt{Genesis} database is part of a larger effort of the Earths in Other Solar Systems (EOS) program aimed at understanding the frequency and formation of earth-like planets with bio-critical ingredients. Such a database will be instrumental in the search for extrasolar planets with atmospheric biosignatures in the Solar neighborhood.

\software{
NumPy \citep{numpy}
SciPy \citep{scipy}
Matplotlib \citep{pyplot}
epos \citep{eposv3}
KeplerPORT \citep{burke15}
}

\acknowledgments
This material is based on work supported by the National
Aeronautics and Space Administration under Agreement No.
NNX15AD94G for the program Earths in Other Solar Systems. The results reported herein benefitted from collaborations and/or information exchange within NASA`s Nexus for Exoplanet System Science (NExSS) research coordination network sponsored by NASA`s Science Mission Directorate
%We acknowledge people. 
We thank Andre Izidoro for providing his modified version of the Mercury integrator.

\bibliographystyle{yahapj}
\bibliography{references,software}

%\appendix
%\section{appendix section}

\end{document}